\documentclass[amsmath,amssymb,amsbsy,twocolumn,prb,showpacs]{revtex4-1}

\def \la {\langle}
\def \ra {\rangle}
\def \ba {\begin{align}}
\def \ea {\end{align}}
\def \ba* {\begin{align}}
\def \ea* {\end{align}}

\usepackage{subfigure}
\usepackage{graphicx}
\usepackage{dcolumn}
\usepackage{bm}
\usepackage[usenames, dvipsnames]{color} 
\usepackage{dsfont} 
\usepackage{footnote}
\usepackage{url}
\usepackage{hyperref}

\begin{document}

\preprint{APS/123-QED}

\title{Family of Sachdev-Ye-Kitaev models motivated by experimental considerations}

\author{\'{E}tienne Lantagne-Hurtubise}
\email{lantagne@phas.ubc.ca}
\author{Chengshu Li}
\email{chengshu@phas.ubc.ca}
\author{Marcel Franz}

\affiliation{Department of Physics and Astronomy \& Quantum Matter Institute, University of British Columbia, Vancouver, British Columbia V6T 1Z1, Canada}


\date{\today}

\begin{abstract}
Several condensed-matter platforms have been proposed recently to realize the Sachdev-Ye-Kitaev  (SYK)  model in their low-energy limit. In these proposed realizations, the characteristic SYK behavior is expected to occur under certain assumptions about the underlying physical system that (i) render all bilinear terms small compared to four-fermion interactions and (ii) ensure that the coupling constants are approximately all-to-all and independent random variables. In this work we explore, both analytically and numerically, the family of models that arises when we relax these assumptions in ways motivated by real physical systems. By relaxing (i) and allowing large bilinear terms, we obtain a novel, exactly-solvable cousin of the SYK model. It exhibits two distinct phases separated by a quantum phase transition characterized by a power-law, $\sim |\omega|^{-1/3}$ scaling of the low-energy spectral density, despite being a non-interacting model. By relaxing (ii), we obtain close relatives of the SYK model which exhibit interesting behaviors, including a chaotic non-Fermi liquid phase with continuously varying fermion scaling dimension, and a phase transition to a disordered Fermi liquid as a function of interaction range and disorder length scale.
\end{abstract}

\maketitle


\section{Introduction}
A model of $N$ Majorana fermions interacting via random all-to-all interactions, the Sachdev-Ye-Kitaev  (SYK) model [\onlinecite{SY1996, Kitaev2015, Maldacena2016}], exhibits intimate connections to the physics of black holes, thermalization and quantum chaos [\onlinecite{Sachdev2015,Maldacena2016b,Hosur2016,Polchinski2016,Verbaar2016}]. The model presents a rare example of an exactly solvable (in the $N \rightarrow \infty$ limit), yet strongly-interacting quantum-mechanical system, and has attracted considerable attention lately as a bridge between several areas of theoretical physics. Its many extensions focus on topics as varied as supersymmetry [\onlinecite{Fu2016,Li2017,Murugan2017,Peng2017}],  quantum chaos [\onlinecite{Chen2017,Krishnan2017}], symmetry-protected topological phases [\onlinecite{You2017,Zhang2018}], higher spatial dimensions [\onlinecite{Berkooz2016,Jian2017,Gu2017,Balents2017, chowdury2018}], strange metals [\onlinecite{Patel2017, Wu2018}], quantum phase transitions [\onlinecite{Altman2016,Bi2017,Zhang2017,garcia2017,garcia2018}], as well as disorder-free versions [\onlinecite{Witten2016}]. Our aim here is to bring these theoretical developments closer to experimental realizations.  
 
The SYK Hamiltonian reads
\begin{align}
H_{\rm SYK} = \sum_{i < j < k < l} J_{ijkl} \chi_i \chi_j \chi_k \chi_l,
\label{hsyk}
\end{align}
where the indices $i,j,k,l$ run over the $N$ Majorana zero-modes (MZMs) in the system, with the usual algebra
\begin{align}
\{ \chi_i, \chi_j \} = \delta_{ij} \quad , 
\quad \chi_i^\dagger = \chi_i.
\label{eq_commutation}
\end{align}
The coupling constants $J_{ijkl}$ are random and independent, and are taken from the Gaussian distribution with
\begin{align}
    \overline{J_{ijkl}} = 0 \quad , \quad \overline{J_{ijkl}^2} = \frac{3!}{N^3} J^2,
\end{align} 
where $J$ is a parameter with dimensions of energy. The SYK model defined in Eq.\ (\ref{hsyk}) is a member of a larger family of SYK$_q$ Hamiltonians with $q$-fermion interactions, where $q$ is an even integer. In this paper, we focus on the SYK$_4$ case already defined in Eq.\ (\ref{hsyk}), and also discuss the SYK$_2$ Hamiltonian.

Several proposals for the physical realization of the SYK model now exist: the Fu-Kane superconductor [\onlinecite{FuKane2008}] threaded with $N$ quanta of magnetic flux in a nanoscale fabricated hole [\onlinecite{Pikulin2017}], and an array of Majorana wires coupled to a quantum dot [\onlinecite{Alicea2017}]. The closely-related complex-fermion Sachdev-Ye (SY) model was argued to occur in cold-atom systems [\onlinecite{Danshita2017}] and in graphene flakes [\onlinecite{Achen2018}] under an applied magnetic field. A proposal for digital quantum simulation of the model has been put forward [\onlinecite{Solano2017}], and first experimental results have been reported [\onlinecite{Luo2017}] consistent with theoretical predictions [\onlinecite{Bi2017}].

The proposals of Refs.~[\onlinecite{Pikulin2017,Alicea2017,Achen2018}] rely on three main assumptions: (i) a symmetry present in the system prohibits  bilinear terms in the Hamiltonian, (ii) random symmetry-preserving disorder on a short length scale introduces the required randomness into four-fermion coupling constants, and (iii) strongly-screened (local) interactions make these approximately Gaussian distributed. In this work, we relax these assumptions in turn and investigate the resulting family of physically-motivated extensions of the canonical SYK model.

The Hamiltonian we study has the form of Eq.\ (\ref{hsyk}), but we consider coupling constants $J_{ijkl}$ with an internal structure that reflects the microscopic properties of the underlying physical system. For short range interactions, they assume a simple form
\begin{equation}\label{JJ}
J_{ijkl} \sim V_0\sum_{m=1}^M \epsilon^{\alpha\beta\mu\nu}\phi_{\alpha i}^m\phi_{\beta j}^m\phi_{\mu k}^m\phi_{\nu l}^m,
\end{equation}  
where $\epsilon^{\alpha\beta\mu\nu}$ denotes the totally antisymmetric tensor,  $\phi_{\alpha i}^m$ with $\alpha=1\cdots 4$ are random independent variables and $V_0$ is the interaction strength. In Sec.\ \ref{sec_Model} we extend this to 
 the screened Coulomb potential with Thomas Fermi length $\lambda$, which results in a slightly more complicated expression for $J_{ijkl}$. We consider both repulsive and attractive interactions and label them by $u=\pm 1$, respectively. The variables $\phi_{\alpha i}^m$ represent coarse-grained components of the single-particle wavefunctions associated with Majorana fermion $\chi_i$ at the spatial point ${\bm r}_m$, and are considered random and independent beyond a disorder length scale $\zeta$. The latter defines $M=L^2/\zeta^2$, where $L$ is the linear size of the two-dimensional region in which Majorana zero modes are confined (see Fig. \ref{fig_chargedensity}). We also introduce the important parameter
\begin{align}
    p = M/N
    \label{eq_p}
\end{align}
which controls the number of terms in the sum Eq.\ (\ref{JJ}) relative to the total number $N$ of zero modes.

As discussed below, the length scales $L$, $\zeta$ and $\lambda$ are a property of the experimental setup and can be controlled by various means. When thinking about the model, it is convenient to fix the disorder scale $\zeta$ and measure $L$ and $\lambda$ relative to it. There are thus two dimensionless parameters that characterize the model, $L/\zeta$ and $\lambda/\zeta$, in addition to $N$. Since we are interested in the large-$N$ limit, it is more convenient to use $p$ and $u\lambda/\zeta$ as two independent parameters describing the model. 

\begin{figure}
\centering
\includegraphics[width = 0.45\textwidth]{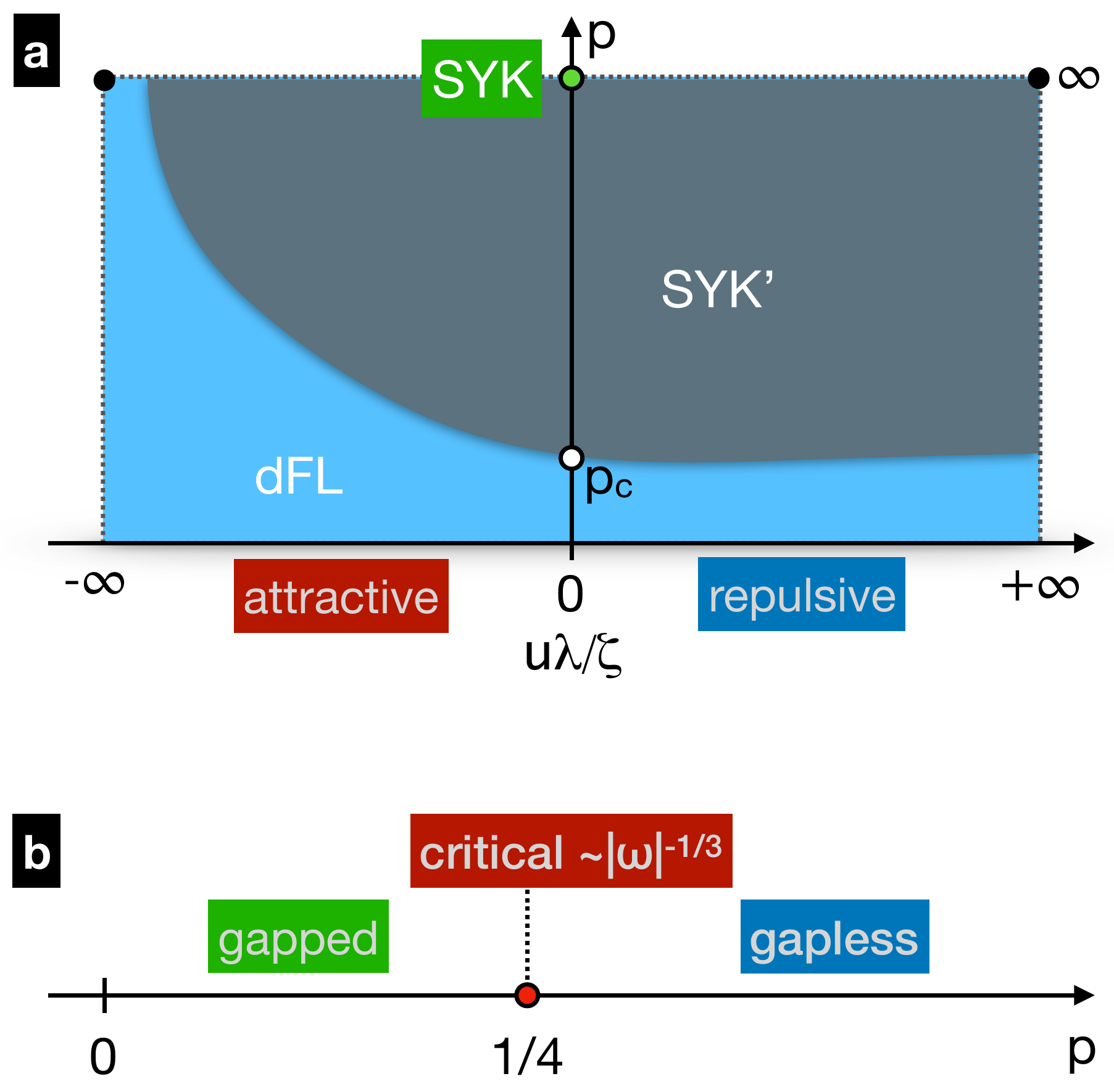}
\caption{a) Conjectured zero-temperature phase diagram of the SYK$_4$ variant discussed in this work. dFL stands for disordered Fermi liquid, and SYK' denotes the SYK-like non-Fermi liquid with continuously varying fermion scaling dimension. $p$, $u$, $\lambda$ and $\zeta$ are model parameters discussed in the text. b) Zero-temperature phase diagram of the SYK$_2$ model variant.}
\label{fig1}
\end{figure}

Our main result is the zero-temperature phase diagram of the extended SYK$_4$ model 
outlined in Fig.\ \ref{fig1}a. It was obtained through a combination of analytical arguments and numerical simulations detailed in the following Sections, focusing on the $\lambda/\zeta=0$ and $p\gg 1$ lines. For large $p$, we find strong evidence for an SYK-like phase, which we denote as SYK' -- a fast-scrambling non-Fermi liquid with continuously varying fermion scaling dimension. At $p\to \infty$ and for short-range interactions (green dot in the phase diagram) the system becomes the maximally chaotic SYK$_4$ model of Refs.\ [\onlinecite{SY1996, Kitaev2015, Maldacena2016}]. At small $p$, we find a disordered Fermi liquid with sharp quasiparticle excitations. For short-range interactions, the quantum phase transition occurs at $p=p_c\simeq {1\over 4}$ as marked by the white dot in the phase diagram.

We also study a variant of the non-interacting SYK$_2$ model in which  matrix elements have internal structure similar to Eq.\ (\ref{JJ}). In this case, we obtain a full analytical solution (in the large-$N$ limit) for the propagator and deduce its phase diagram as a function of parameter $p$ (Fig.\ \ref{fig1}b). A quantum phase transition from a gapped phase at $p<p_c={1\over 4}$ to a gapless phase at $p>p_c$ occurs, and the critical point is characterized by a scale-invariant propagator $\sim|\omega|^{-1/3}$ at low energies.  

The rest of this paper is organized as follows. In Sec.~\ref{sec_Model}, we briefly review physical realizations of SYK physics and define our model. In Sec.~\ref{sec_H2}, we explore the physics away from the exact symmetry point prohibiting bilinear terms, and solve the corresponding non-interacting model exactly. In Sec.~\ref{sec_p}, we analyze the effect of varying disorder length scale, while in Sec.~\ref{sec_lambda} we consider a screened Coulomb potential with arbitrary sign and range. In all cases, we perform extensive numerical simulations and extract spectral and thermodynamic quantities, Green's functions, and out-of-time-order correlators. Finally, Sec.~\ref{sec_Dis} concludes with a summary and open questions. Various details are relegated to the appendices.

\section{The Model}
\label{sec_Model}

A common feature of the solid-state realizations of the SYK model discussed in Refs. [\onlinecite{Pikulin2017,Alicea2017,Achen2018}] is that the randomness in coupling constants $J_{ijkl}$ arises from the random spatial structure of the wavefunctions $\Phi_j({\bf r})$ describing the active fermionic degrees of freedom. This random structure arises in turn from the microscopic disorder present in the system. The key requirement, which makes the system non-generic, is that the microscopic disorder respects a symmetry that prohibits the formation of fermionic bilinears in the Hamiltonian. 

\begin{figure}
\centering
\includegraphics[width = 0.42\textwidth]{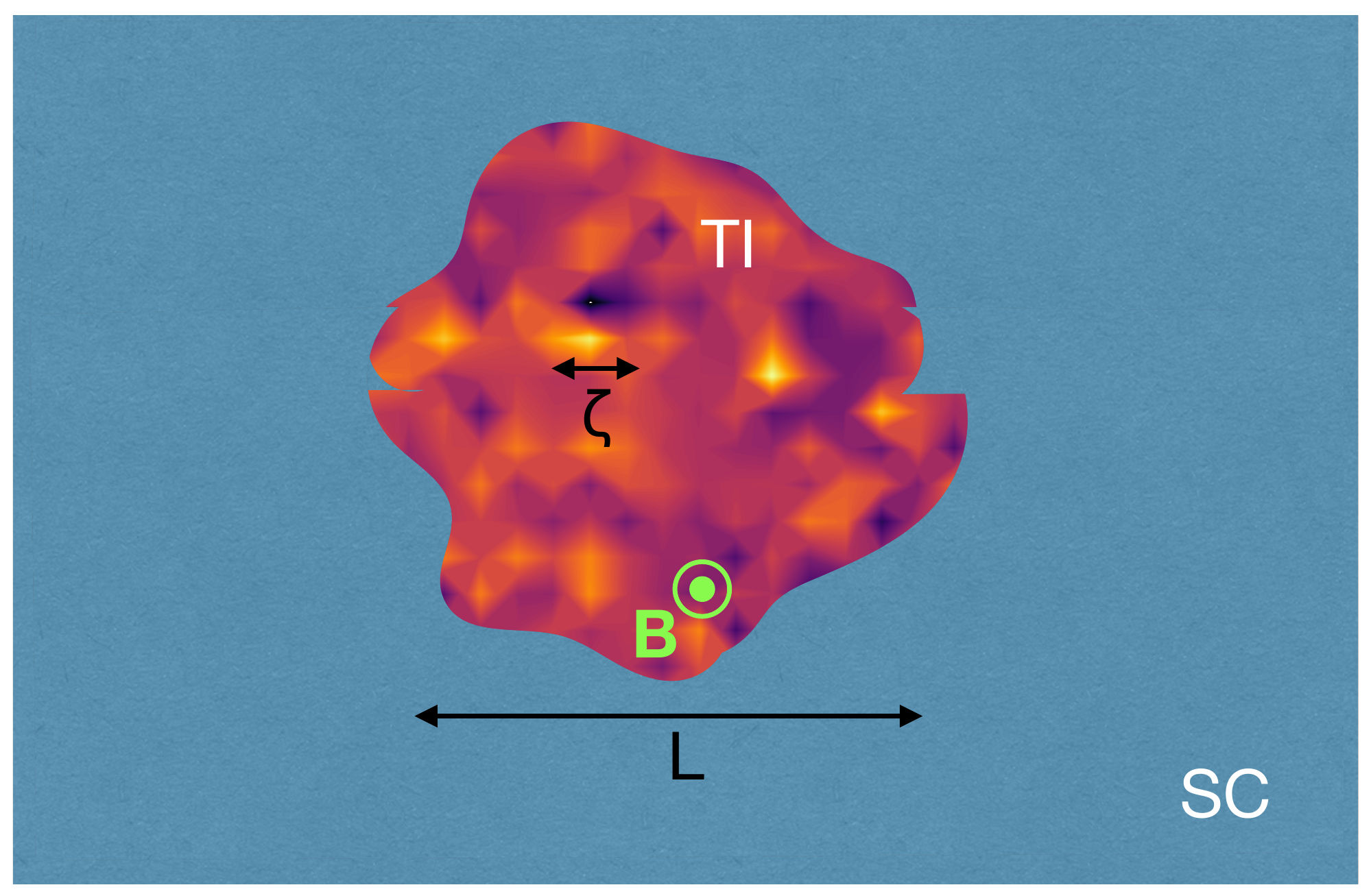}
\caption{Schematic depiction of the device proposed in Ref.\ [\onlinecite{Pikulin2017}]: a 3D topological insulator (TI) covered by a thin layer of an ordinary superconductor (SC). A hole with an irregular shape is fabricated in the SC layer and threaded with magnetic flux $\Phi$. The colored region represents the hole of linear size $L$. The colorscale inside the hole illustrates the typical charge density distribution associated with a pair of MZMs  with the characteristic disorder length scale $\zeta$.}
\label{fig_chargedensity}
\end{figure}

For example, Ref.\ [\onlinecite{Pikulin2017}] employs an interface between a 3D topological insulator (TI) and an ordinary superconductor (SC), as illustrated in Fig.\ \ref{fig_chargedensity}. It is known that vortices in the topological Fu-Kane superconductor formed by such an interface host Majorana zero modes [\onlinecite{FuKane2008}]. If a nanoscale hole is fabricated in the SC and threaded with $N$ magnetic flux quanta, the same number $N$ of MZMs are bound to the hole. At the neutrality point (when the chemical potential $\mu$ of the TI coincides with the surface state Dirac point), the MZMs are protected by the fictitious time reversal symmetry $\Sigma$ and remain at exactly zero energy [\onlinecite{TeoKane2010}]. As a result, no fermion bilinears are permitted in the Hamiltonian describing the zero modes. Crucially, a hole with an irregular shape preserves $\Sigma$ but imparts random spatial structure onto the MZM wavefunctions. This leads to random coupling constants $J_{ijkl}$ when short-range Coulomb interactions are included and the low-energy physics is therefore described by Eqs.~(\ref{hsyk}) and (\ref{JJ}).

We thus consider $N$ MZMs confined in a spatial region of size $\sim L$. We make a key simplifying assumption, rooted in investigations of realistic model systems: that wavefunctions $\Phi_j({\bf r})$ corresponding to the different MZMs are well approximated by uncorrelated, randomly-distributed functions over a characteristic length scale $\zeta$. A screened Coulomb potential with arbitrary range $\lambda$ and sign is then applied to the system. This is similar in spirit to the setup of Ref.~[\onlinecite{Pikulin2017}] -- however, here we consider an idealized case where the randomness is introduced directly into the wavefunctions and is thus fully controllable. The setup in Refs.\ [\onlinecite{Alicea2017}] and [\onlinecite{Achen2018}] differs in details but the underlying mechanism for generating random coupling constants is similar. In the following, we focus our discussion on the architecture of Ref.~[\onlinecite{Pikulin2017}] but expect our results to be broadly relevant to all solid-state realizations of the SYK and SY models.

\subsection{Setup}

The wavefunctions of the $N$ MZMs in Ref.~[\onlinecite{Pikulin2017}]  are 4-component complex spinors $\Phi_i(\mathbf{r})$ in the combined spin and Nambu space. The reality condition on the MZMs,
\begin{align} \sigma^y \tau^y \Phi_j(\mathbf{r})^* = \Phi_j(\mathbf{r}),
\end{align}
where ${\bm \sigma}$ and ${\bm \tau}$ are Pauli matrices in spin and Nambu space, respectively, leads to
\begin{align}
    \Phi_j(\mathbf{r}) = 
\begin{pmatrix}
\eta_j(\mathbf{r}) \\
i \sigma^y \eta_j^*(\mathbf{r})
\end{pmatrix},
\end{align} 
where $\eta_j(\mathbf{r})$ are two-component complex spinors. We consider the MZM wavefunctions to be randomly distributed both in real space (over a characteristic length scale $\zeta$) and in spin space. To implement this random structure, we discretize the wavefunctions on a lattice with spacing $\zeta$ and $M \sim L^2/\zeta^2$ lattice points, writing
\begin{align}
 \eta_j(\mathbf{r}_m) \equiv \begin{pmatrix}
\phi_{j1}^m + i \phi_{j2}^m \\
\phi_{j3}^m + i \phi_{j4}^m
\end{pmatrix}     .
\end{align}
The discretized spinors are assumed to be composed of real, uncorrelated random Gaussian variables $\phi_{j\alpha}^m$ [$\alpha = 1..4$, $j=1..N$,  $m = 1.. M$]. Each spinor is normalized, $1 = \sum_m |\Phi_j(\mathbf{r}_m)|^2$, which specifies the variance of the random variables $\phi_{j\alpha}^m$ as 
\begin{align}
\overline{\left( \phi_{j\alpha}^m  \right)^2} = \frac{1}{8M}.
\label{var0}
\end{align}
The wavefunctions are orthonormal \textit{on average},
\begin{align}
\delta_{ij} = \sum_m  \overline{\Phi_i^\dagger(\mathbf{r}_m) \Phi_j(\mathbf{r}_m)}
= 2\sum_{m, \alpha} \overline{\phi_{i\alpha}^m \phi_{j\alpha}^m} \quad \forall \; i, j
\label{eq_orthogonality_condition}
\end{align}
because the $\phi_{i\alpha}^m$ are uncorrelated and have zero mean.
We can quantify the deviations from exact orthonormality by calculating the variance
\begin{align}
\mathrm{Var}\big(2\sum_{m,\alpha}\phi_{i,\alpha}^m\phi_{j,\alpha}^m\big)=\frac{1+\delta_{ij}}{4M}.
\end{align}
Thus the orthonormality becomes exact as $M\rightarrow\infty$.

\subsection{Interactions}

The setup described above can be thought of, more generally, as the idealized low-energy subspace of a non-interacting Hamiltonian with the appropriate symmetries for the topological class BDI (protecting $N$ MZMs) and strong random disorder consistent with those symmetries. Specifically, it is crucial that an antiunitary symmetry prohibits the formation of bilinear terms, so that the four-fermion interaction terms become dominant. This situation is described by a generic Hamiltonian
\begin{align} 
H_4 = \sum_{i<j<k<l} J_{ijkl} \chi_i \chi_j \chi_k \chi_l. 
\label{eq_H}
\end{align}
For density-density interactions between the underlying electrons, Ref.\ [\onlinecite{Pikulin2017}] showed that 
\begin{align}
J_{ijkl} = \frac{1}{3} \left( \tilde{J}_{ijkl} + \tilde{J}_{iljk} - \tilde{J}_{ikjl} \right)
\label{eq_antisymmetrization}
\end{align}
where
\begin{align}
\tilde{J}_{ijkl} = - \sum_{m,n} \rho_{ij}^m V_{mn} \rho_{kl}^n  .
\label{eq_couplings}
\end{align}
Here, $V_{mn}$ is the interaction potential and
\begin{align}\rho_{ij}^m = \frac{i}{2} \Phi_i^\dagger(\mathbf{r}_m) \tau^z \Phi_j(\mathbf{r}_m) =  -\text{Im}\left[ \eta_i^\dagger(\mathbf{r}_m) \eta_j(\mathbf{r}_m) \right] \end{align}
is the charge density associated with the pair of MZMs $\chi_i$, $\chi_j$. The charge density is real and anti-symmetric in $i$,$j$ (see Fig. \ref{fig_chargedensity} for a representative example). Only the totally anti-symmetric part of $\tilde{J}_{ijkl}$ contributes to Eq. (\ref{eq_H}), as enforced by Eq. (\ref{eq_antisymmetrization}). In the following, we assume a screened Coulomb potential
\begin{align}
 V_{mn} = V_0 \frac{ e^{-r_{mn}/\lambda} }{\sqrt{r^2_{mn} + \zeta^2}}  \quad , \quad V_0 \equiv u \frac{
 2 \pi e^2}{\epsilon}
\label{eq_Coulomb}
\end{align} 
where $\lambda$ is the Thomas-Fermi screening length, $\epsilon$ is the dielectric constant, $u = \pm 1$ and we have regularized the potential for small distances, i.e. $V_{mm} = V_0/\zeta$. For Coulomb repulsion, the physical sign is $u=+1$ but in the following we consider either sign. Indeed, MZMs occur in superconductors where attractive electron-electron interaction is generically required to form Cooper pairs. The bulk of such attraction is subsumed into the mean-field BCS treatment, but there are always residual interactions which can, presumably, be both attractive and repulsive depending on the details.

\subsection{Relevant length scales and limits}

Our model possesses three relevant length scales: $\zeta$ controls the scale of spatial disorder (and thus, our ``lattice spacing"),
 $L$ defines the size of the region where the MZMs are localized, and $\lambda$ determines the range of Coulomb interactions. 
One expects to obtain the SYK physics in the limit $\lambda \ll \zeta \ll L$ (strongly-screened interactions and large hole). Intuitively, this becomes clear noting that the interactions are essentially on-site ($ V_{mn} = \frac{V_0}{\zeta} \delta_{mn} $), leading to
\begin{align}
J_{ijkl} = - \frac{V_0}{3 \zeta} \sum_{m=1}^{M} \epsilon^{\alpha \beta \mu \nu} \phi_{i \alpha}^m \phi_{j\beta}^m \phi_{k\mu}^m \phi_{l\nu}^m,
\label{on-site-J}
\end{align} 
and the number of sites $M \sim L^2/\zeta^2  \gg N$. The coupling constants are now given by a sum over a large number of identically-distributed random variables, and become asymptotically Gaussian by virtue of the central limit theorem. This argument will be supported by a field theoretic calculation in Appendix \ref{App-4phi}. 

We thus have three ``knobs" to turn to explore the physics away from the SYK limit. (1) Tuning the chemical potential $\mu \neq 0$ generates bilinear terms which ultimately destroy the non-Fermi liquid at low energies, but the resulting physics is nevertheless interesting (Sec.~\ref{sec_H2}). (2) Decreasing the size of the hole, and thus reducing the number of terms $M$ in the sum, allows us to move away from the Gaussian distribution enforced by the central limit theorem (Sec.~\ref{sec_p}). Increasing the magnetic field leads to larger $N$ and has the same effect. (3) Finally, increasing $\lambda$ changes the statistical distribution of $J_{ijkl}$ away from Gaussian (Sec.~\ref{sec_lambda}).

In the following Sections we explore the effects of the perturbations mentioned above.

\section{SYK$_2$ model variant}
\label{sec_H2}

In order to access the strongly-interacting regime of the SYK$_4$ model, it is crucial that bilinear terms are forbidden from the low-energy physics by a  symmetry. For the Fu-Kane model realization discussed in Ref.\ [\onlinecite{Pikulin2017}], this is the fictitious time reversal $\Sigma$  present at the $\mu = 0$ neutrality point. Indeed, simple scaling arguments show that such bilinear terms, if present, will dominate the low-energy physics and destroy the non-Fermi liquid SYK state. In this section, we provide further insights into the peculiar Fermi liquid state obtained when the chemical potential is tuned far from neutrality point, i.e. $\mu \gg J$. As we shall see, the resulting model is exactly solvable in the large-$N$ limit and it exhibits interesting spectral properties that depend nontrivially on the parameter $p$ defined in Eq.\ (\ref{eq_p}) [see phase diagram in Fig. \ref{fig1}b]. This solution also provides guidance for the fully interacting model which we consider in the following Sections.
\begin{figure*}
\centering
\includegraphics[width = 0.95\textwidth]{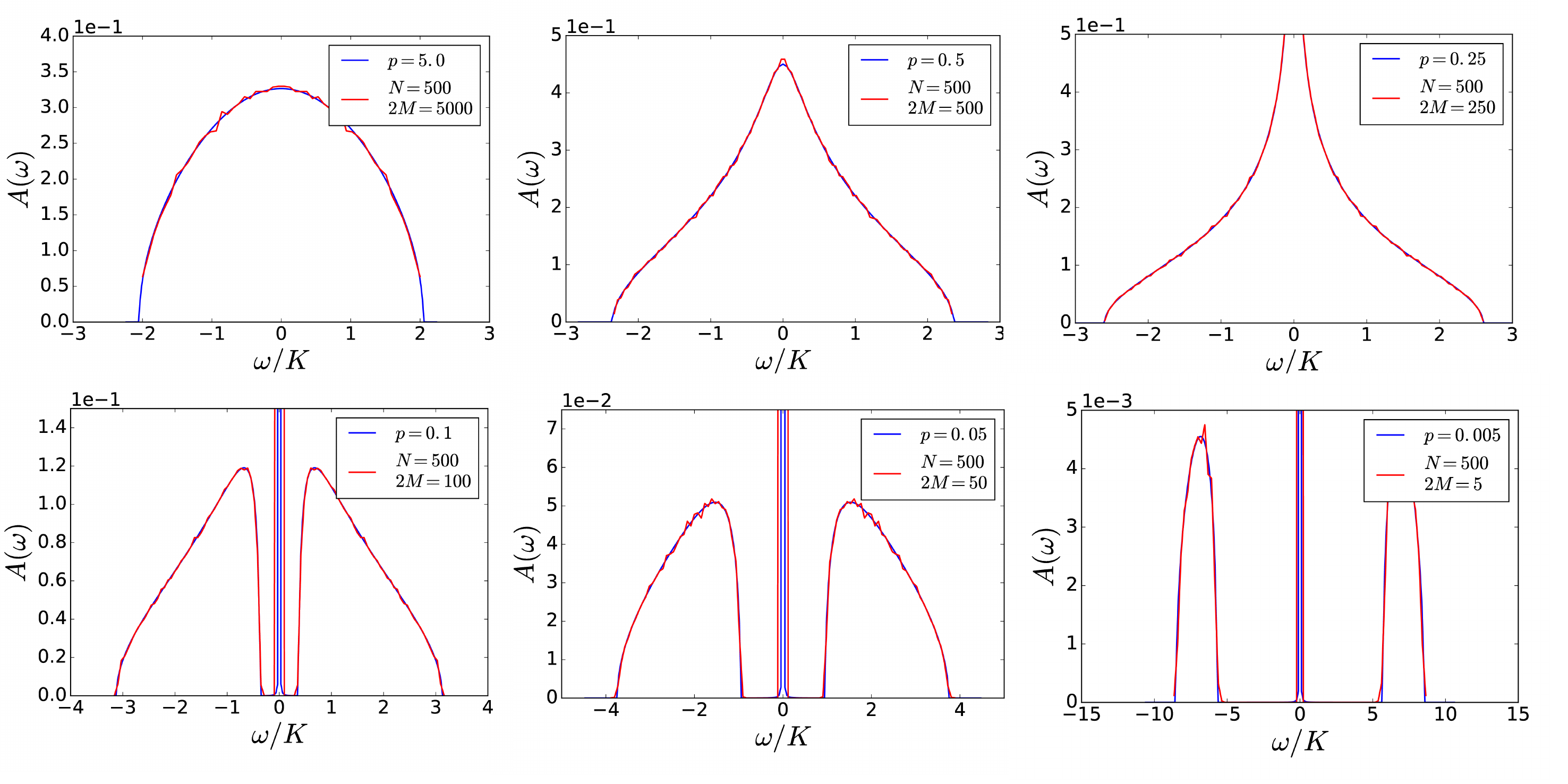}
\caption{Comparison between the spectral function of the modified SYK$_2$ model obtained from the large-$N$ calculation (blue) and from exact diagonalization (red), averaged over many disorder realizations.}
\label{fig_non_int}
\end{figure*}

For $\mu \gg J$, we can neglect the four-fermion terms and the Hamiltonian takes the form
\begin{align} 
H_2 = i \sum_{i<j} K_{ij} \chi_i \chi_j,
\label{eq_H2}
\end{align}
where the ``hopping" constants are given by
\begin{align}
K_{ij} =& -\mu \sum_{m=1}^M \rho_{ij}^m
= -\mu\sum_{m=1}^{2M} \left( a_i^m b_j^m - b_i^m a_j^m\right),
\label{eq_couplings_H2}
\end{align}
and we defined $(a_j^m,a_j^{m+M})=(\phi_{j1}^m,\phi_{j3}^m)$ and  $(b_j^m,b_j^{m+M})=(\phi_{j2}^m,\phi_{j4}^m)$.
As before, the wavefunction components are taken to be random independent Gaussian variables. Specifically, Eq.\ (\ref{var0}) implies 
\begin{align}
\overline{a_i^m a_j^n}=\overline{b_i^m b_j^n}={1\over 8M}\delta_{ij}\delta^{mn}, \ \ \ \overline{a_i^m b_j^n}=0.
\label{varab}
\end{align}
Viewed as $N$-component vectors in indices $i,j$ we see that ${\bf a}^m$ and ${\bf b}^m$ are approximately orthogonal. Clearly, this can hold only when $4 M < N$ because at most $N$ $N$-component vectors can be mutually orthogonal. 

The problem defined by Hamiltonian (\ref{eq_H2}) can be solved by diagonalizing the $N\times N$ Hermitian matrix $iK_{ij}$. Assuming that all vectors ${\bf a}^m$ and ${\bf b}^m$ are mutually orthogonal (exactly, not only on average), it is straightforward to show that the matrix $iK_{ij}$ has two types of eigenvectors when $4 M< N$. It has $4M$ eigenvectors $({\bf a}^m\pm i{\bf b}^m)/\sqrt{2}$ with eigenvalues $\pm\mu/8p$ and $(N-4M)$ eigenvectors orthogonal to ${\bf a}^m$ and ${\bf b}^m$ with zero eigenvalues. Therefore, the spectral function of $H_2$ consists of a $\delta$-function peak at zero energy associated with the zero modes and two satellite peaks at  $\pm\mu/8p$. The ground state of the system is gapped and exhibits degeneracy $2^{(N-4M)/2}$ owing to the Majorana zero modes.

When the exact orthonormality is relaxed (but continues to hold on average), we expect this picture to remain valid as long as $4 M \ll N$. The spectral peaks may broaden due to fluctuations in vectors ${\bf a}^m$ and ${\bf b}^m$ away from perfect orthonormality. On the other hand, we expect the picture to break down when $4M$ approaches $N$ because, at that point, the vectors can no longer be mutually orthogonal. A phase transition to a gapless state is therefore expected at or near $p=1/4$. Indeed, in the opposite limit $p\gg 1/4$ (or $4M\gg N$), the sum in Eq.\ (\ref{eq_couplings_H2}) contains many identically distributed random terms and $i K_{ij}$ becomes Gaussian distributed due to the central limit theorem. In this situation, we expect the spectral function to approach the semicircle law, 
\begin{align}
A(\omega)={1\over \pi K}{\rm Re}\sqrt{1-\left({\omega\over 2 K}\right)^2},
\label{semi}
\end{align}
where 
\begin{align}
K^2\equiv  N\overline{K_{ij}^2}=\mu^2/{16 p},
\label{K}
\end{align}
and the last equality follows from Eqs.\ (\ref{eq_couplings_H2}) and (\ref{varab}).

The model defined by Eqs.\ (\ref{eq_H2}-\ref{varab}) can be solved analytically in the large-$N$ limit by functional integral techniques used to average over disorder realizations.  The result of this procedure, presented in Appendix \ref{app_H2}, is a single cubic equation for the averaged Green's function $G(\tau) \equiv \frac{1}{N}\sum_i\langle T_\tau \chi_i(\tau)\chi_i(0)\rangle$ in the Matsubara frequency space,
\begin{align}
 i \omega K^2 G^3 + (1 - 4p)K^2 G^2 - 4 i\omega p G - 4p  = 0.
\label{eq_H2_selfconsistent_1}
\end{align}
We observe that the parameter $p = M/N$ affects non-trivially the form of the solution. Specifically, as anticipated from our general argument, at $p=1/4$
the prefactor of the second term changes sign, signaling a phase transition. The low-energy scaling of the spectral function at this critical point is $\sim |\omega|^{-1/3}$. This power-law scaling is perhaps surprising, as our model is ultimately non-interacting, with well-defined quasiparticle excitations. It can also be checked that, for small $p$, the solution of Eq.\ (\ref{eq_H2_selfconsistent_1}) gives the low-frequency spectral function $A(\omega)\simeq(1-4p)\delta(\omega)$, in agreement with our heuristic analysis implying $N(1-4p)$ zero modes, while for large $p$ it approaches the semicircle law Eq.\ (\ref{semi}).

In Fig.\ \ref{fig_non_int}, we compare the numerical solution of  Eq.~(\ref{eq_H2_selfconsistent_1}) with the result of an exact diagonalization of the Hamiltonian Eq.~(\ref{eq_H2}). 
For large values of $p$, we recover the expected semicircle distribution. As $p$ is lowered, a low-energy peak starts developing. At the critical point, $p_c=1/4$, we find the expected $\sim |\omega|^{-1/3}$ divergence at low energies. Further decrease in $p$ below $1/4$ reveals a gapped spectrum with excitations given by two symmetric lobes, centered around $\omega=\pm \mu/8p = \pm K/2\sqrt{p}$. The sharp peak at $\omega=0$ represents the expected zero modes whose spectral weight is $(1-4p)$.

\section{SYK$_4$ variant I: short-range interaction potential}
\label{sec_p}

In this section, we explore the effect of the ratio $p=M/N$ on the fully interacting model. In the physical realization of Ref.\ [\onlinecite{Pikulin2017}], $p$ can be controlled by varying the applied magnetic field or the size of the hole in the superconductor used to trap the magnetic flux. For simplicity, we assume on-site interactions [Eq. (\ref{on-site-J})]. We are thus exploring the vertical axis at $u\lambda/\zeta=0$ in the phase diagram (Fig.\ 1). The Hamiltonian in this case reads
\begin{align}\label{h1}
H_4=-\sum_{i<j<k<l} \sum_{m=1}^M\epsilon^{\alpha\beta\mu\nu}\phi_{i\alpha}^m\phi_{j\beta}^m\phi_{k\mu}^m\phi_{l\nu}^m\chi_i\chi_j\chi_k\chi_l
\end{align}
where we set $V_0/3\zeta=1$ for convenience. 
Note that this model is similar to the one introduced in Ref. [\onlinecite{Bi2017}], but presents a greater analytical challenge, as the coupling constants are a product of \emph{four} independent random variables instead of two.

\subsection{Limiting cases}
The Hamiltonian (\ref{h1}) can be written as 
\begin{equation}
H_4=\frac{1}{2}\sum_m h_m^2    
\end{equation}
with 
\begin{align}
h_m=i\sum_{i< j} \left(\phi_{i,1}^m\phi_{j,2}^m + \phi_{i,3}^m\phi_{j,4}^m - \left(i \leftrightarrow j\right)  \right)\chi_i\chi_j,
\end{align}
representing a sum over squares of Hamiltonians $h_m$ describing non-interacting systems. These have the same structure as the Hamiltonian $H_2$ studied in the previous Section. Importantly,  because of the approximate orthonormality of the wavefunctions $\phi_{j,\alpha}^m$ expressed in Eq.\ (\ref{eq_orthogonality_condition}), one can show that different $h_m$ approximately commute with one another, $[h_m,h_n]\approx 0$, as long as $p\ll 1$. For small $p$, $H_4$ is therefore composed of a sum of approximately commuting operators $h_m^2$ whose individual eigenstates can be understood easily on the basis of our analysis in the previous Section. If $|\Psi_E^m\rangle$ is a many-body eigenstate of $h_m$ with energy $E$, then clearly it is also an eigenstate of $h_m^2$ with energy $E^2$. Therefore, the ground state and low-lying excited states of $h_m^2$ are simply those eigenstates of $h_m$ with energies $E$ closest to zero. Such a ground state is clearly a disordered Fermi liquid (dFL) whose excitations are sharp quasiparticles. Because individual $h_m$ are approximately commuting, we expect the ground state of $H_4$ to also be a dFL in this limit.

For $p\gg 1$, we already argued that each coefficient multiplying $\chi_i\chi_j\chi_k\chi_l$ in Eq.\ (\ref{h1}) is a sum of many identically distributed random variables and therefore tends to a Gaussian distributed random variable. In this limit, we expect the model to approach the SYK behavior. In Appendix~\ref{App-4phi}, we perform 
 a large-$N$ saddle-point calculation to confirm this conjecture. We find that the saddle-point equations for $H_4$ coincide, to lowest order, with those describing the SYK model at large $N$. 
\begin{figure*}
\centering
\includegraphics[width = 0.95\textwidth]{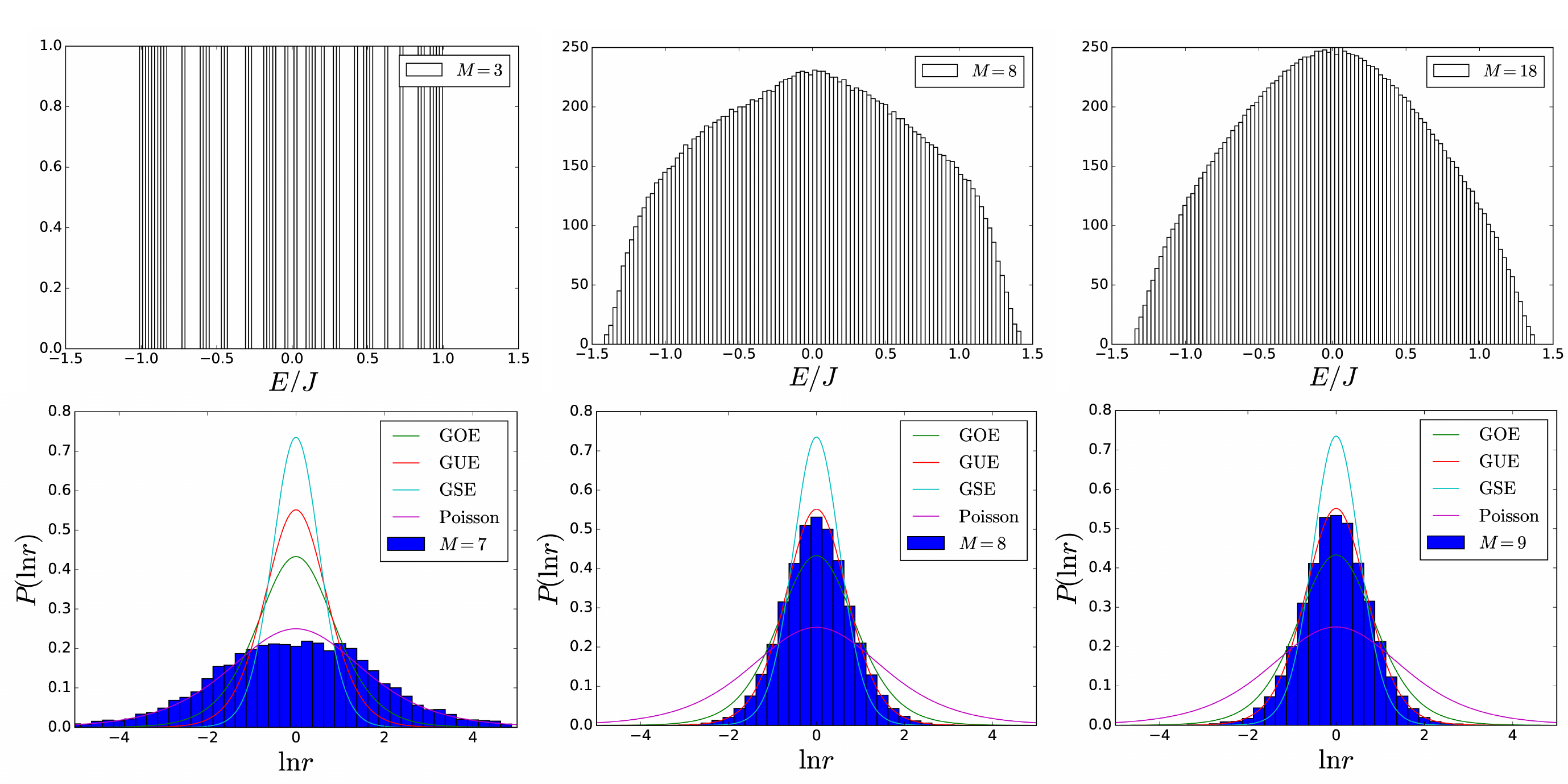}
\caption{Energy eigenvalues (top panels) and level statistics (bottom panels) for various values of $M$ with $N=30$ for short-range interactions. 
}
\label{E_stat_M}
\end{figure*}

 Given the two limiting behaviors discussed above, we expect a phase transition to occur in the model as a function of $p$ from a dFL to a maximally chaotic non-Fermi liquid. As in Sec.\ III, we expect the transition to be driven by an ``orthogonality catastrophe" of sorts: when $4M$ approaches $N$, the wavefunctions $\phi_{j,\alpha}^m$, viewed as $N$-component vectors in index $j$, can no longer be mutually orthogonal because the number of vectors exceeds the size of the vector space. At that point, our arguments pointing to the dFL ground state fail and a phase transition is expected at or near $p={1\over 4}$.    
 In subsequent subsections, we present numerical results in support of this scenario. 

\subsection{Many-body spectra and level statistics}
\begin{center}
\begin{table}
\begin{tabular*}{0.45\textwidth}{@{\extracolsep{\fill}} c | c c c c }
  \hline			
  $N$ (mod 8) & 0 & 2 & 4 & 6 \\
  \hline
  Gaussian ensemble  & GOE & GUE & GSE & GUE \\
  $\beta$  & 1 & 2 & 4 & 2 \\
  $Z$      & $\frac{8}{27}$ & $\frac{4 \pi}{81\sqrt{3}}$ & $\frac{4 \pi}{729\sqrt{3}}$ & $\frac{4 \pi}{81\sqrt{3}}$ \\
\vspace{-6pt}\\
  \hline
\end{tabular*}
\caption{Gaussian ensembles for even $N$}
\label{tab_statistics}
\end{table}
\end{center}
The many-body energy spectrum and the level statistics constitute a useful diagnostic of SYK  physics [\onlinecite{Verbaar2016,You2017}].
The level statistics analysis requires arranging the many-body eigenenergies $E_n$ in ascending order: $E_1 < E_2 < ... < E_{2^{N/2}}$, and forming ratios between successive energy spacings
\begin{align}\label{rn}
r_n = \frac{E_{n+1} - E_n}{E_{n+2}-E_{n+1}}.
\end{align}
The probability distribution of the $\{ r_n \}$ is predicted by random matrix theory to cycle through the ``Wigner surmise'' functions:
\begin{align}
P(r) = \frac{1}{Z} \frac{(r+r^2)^\beta}{(1+r+r^2)^{1+3\beta/2}} 
\end{align}
corresponding to Wigner-Dyson random matrix ensembles with a $Z_8$ periodicity in the number of Majorana modes $N$. The parameters $\beta$ and $Z$ are fixed in any given ensemble, as shown in Table \ref{tab_statistics} -- therefore, this analysis does not require \textit{any} free parameters. As emphasized in Ref.\  [\onlinecite{You2017}], the SYK model conserves fermion parity and the level statistics analysis must be carried out in each parity sector separately. 

Typical results for various $M$ are shown in Fig.~\ref{E_stat_M}, where we take $N=30$. As $M$ increases, the energy distribution evolves from being highly degenerate, with a discrete spectrum, to assuming a continuous distribution characteristic of the SYK physics. For small and large $M$ the level statistics agree with the Poisson and GUE distributions, respectively, lending support for a Fermi liquid phase and an SYK-like phase in the two limits. Numerically, we find that the transition occurs between $M=7$ and $8$, which for $N=30$ corresponds to the two values of $p$  closest to the conjectured critical point $p_c={1\over 4}$.

\subsection{Green's functions}

Another useful diagnostic is obtained through calculating the two-point Green's functions of the model.
This requires full diagonalization of the many-body Hamiltonian (\ref{h1}) to obtain the complete set of eigenvalues and eigenvectors. Using the Lehman representation of many-body eigenstates $| n \ra$ with eigenenergies $E_n$, the retarded Green's function reads
\begin{align}
G_i^R(\omega) = \sum_n \left[ \frac{|\la n|\chi_i|0\ra |^2}{\omega +E_0 - E_n + i\delta} + \left(E_0 \leftrightarrow E_n\right) \right],
\end{align}
where $\delta > 0$ is a small real number. The averaged spectral function, given by
\begin{align}
A(\omega) = -\frac{1}{N \pi} \sum_{i=1}^N  \text{Im} G_i^R(\omega),
\label{eq_Aomega}
\end{align}
would correspond, in a tunneling experiment, to the tunneling conductance measured by a large probe coupling to all modes in the system.

In the SYK phase, the imaginary (Matsubara) frequency Green's functions show a power-law behavior in the conformal limit $\frac{J}{N} \ll \omega \ll J$, 
\begin{align}
-i G(\omega_m) \sim 1/\omega_m^\alpha ,
\label{eq_GMastubara}
\end{align}
with an exponent $\alpha = 1 - 2 \Delta_F$, where $\Delta_F$ is the so-called scaling dimension of the fermion operator in the CFT. For the SYK model, the fermion scaling dimension is $\Delta_F = 1/4$ and thus $\alpha={1\over 2}$. The Matsubara Green's function is obtained from its real-frequency counterpart by analytical continuation, $G(\omega_m) = G^R(\omega +i\delta\rightarrow -i \omega_m )$.

\begin{figure*}
\centering
\includegraphics[width=0.95\textwidth]{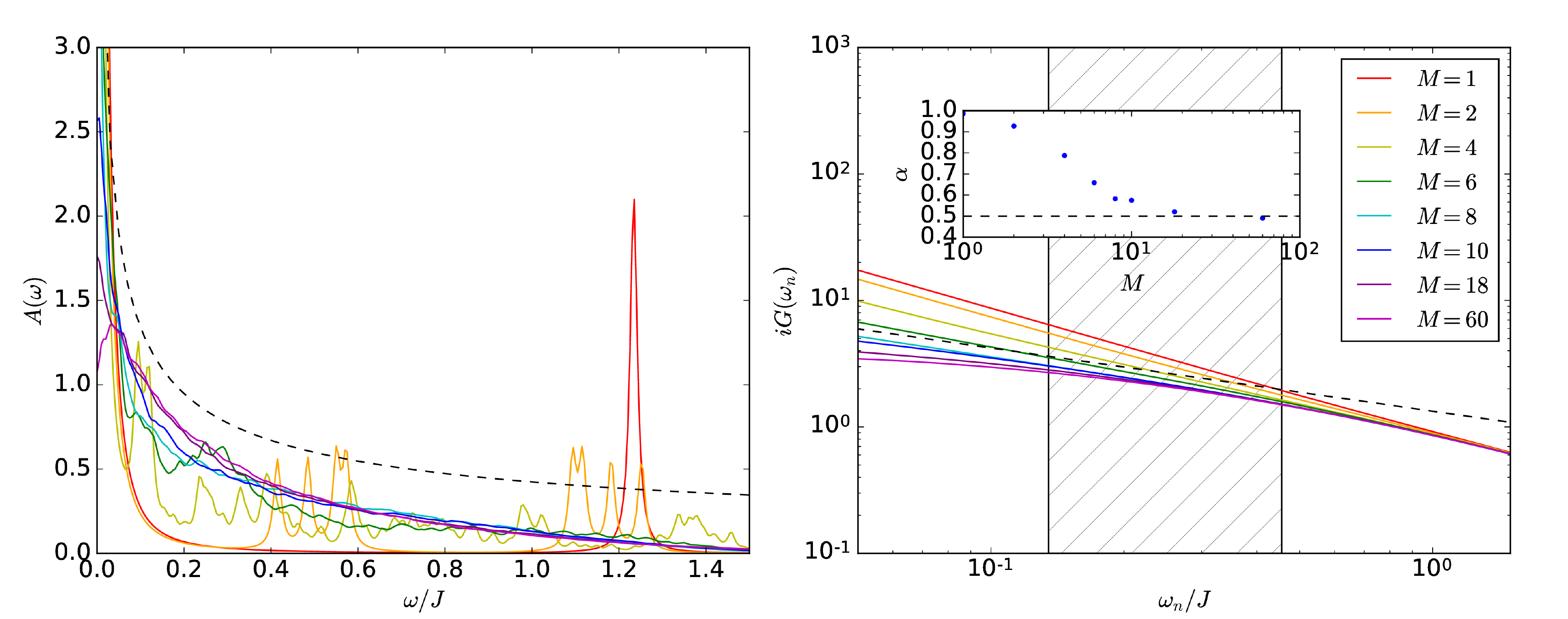}
\caption{Spectral function $A(\omega)$ (left) and Matsubara Green's function $G(\omega_n)$ (right) for $N=30$ with various $M$. Dashed lines represent the $1/\sqrt{\omega}$ behavior expected on the basis of the large-$N$ theory. The inset shows the corresponding exponents extracted from the power-law fit in the shaded window.}
\label{fig_p_Greens}
\end{figure*}

Fig.\ \ref{fig_p_Greens} shows our numerical results for $N=30$. 
The distinction between the two phases (dFL and SYK) is most apparent from the spectral function. For large $p$, $A(\omega)\sim |\omega|^{-1/2}$ at low frequency, while for small $p$ discrete peaks emerge representing the quasiparticle excitations. From the Matsubara Green's function, one sees a continuous change between the two phases, with the exponent $\alpha$ more easily extractable. Fig.\ \ref{fig_p_Greens} (right) shows the window used for the power law fit of the Matsubara Green's function and the extracted exponent $\alpha$, which varies from ${1\over 2}$ at large $p$, characteristic of the SYK conformal scaling, to 1 at small $p$, as expected of a dFL. 

\subsection{Out-of-time-order correlators and scrambling}

Another fascinating aspect of the SYK model is that it maximizes scrambling of quantum information, and is thus maximally chaotic, in analogy to a black hole. Here we numerically obtain a useful measure of scrambling, the out-of-time-order correlator (OTOC), defined as
\begin{align}
F_{ij}(t) = \la \chi_i(t) \chi_j(0) \chi_i(t) \chi_j(0) \ra.
\label{eq_OTOC}
\end{align}
We expect $F_{ij}(t) = 1 - (J/TN) \exp (\lambda_L t) $ in the conformal regime, during the scrambling process $\beta \ll t < \beta \log N$ [\onlinecite{Kitaev2015,Maldacena2016}]. This is however difficult to access in numerical calculations [\onlinecite{FuSachdev2016}] because of the smallness of $N$.
The exponent $\lambda_L$, an analog of the Lyapunov exponent in classical chaos, is expected to saturate the universal bound $\lambda_L = 2 \pi T$ conjectured in Ref.~[\onlinecite{Maldacena2016b}] based on general AdS/CFT duality arguments.

\begin{figure}
\centering
\includegraphics[width = 0.45\textwidth]{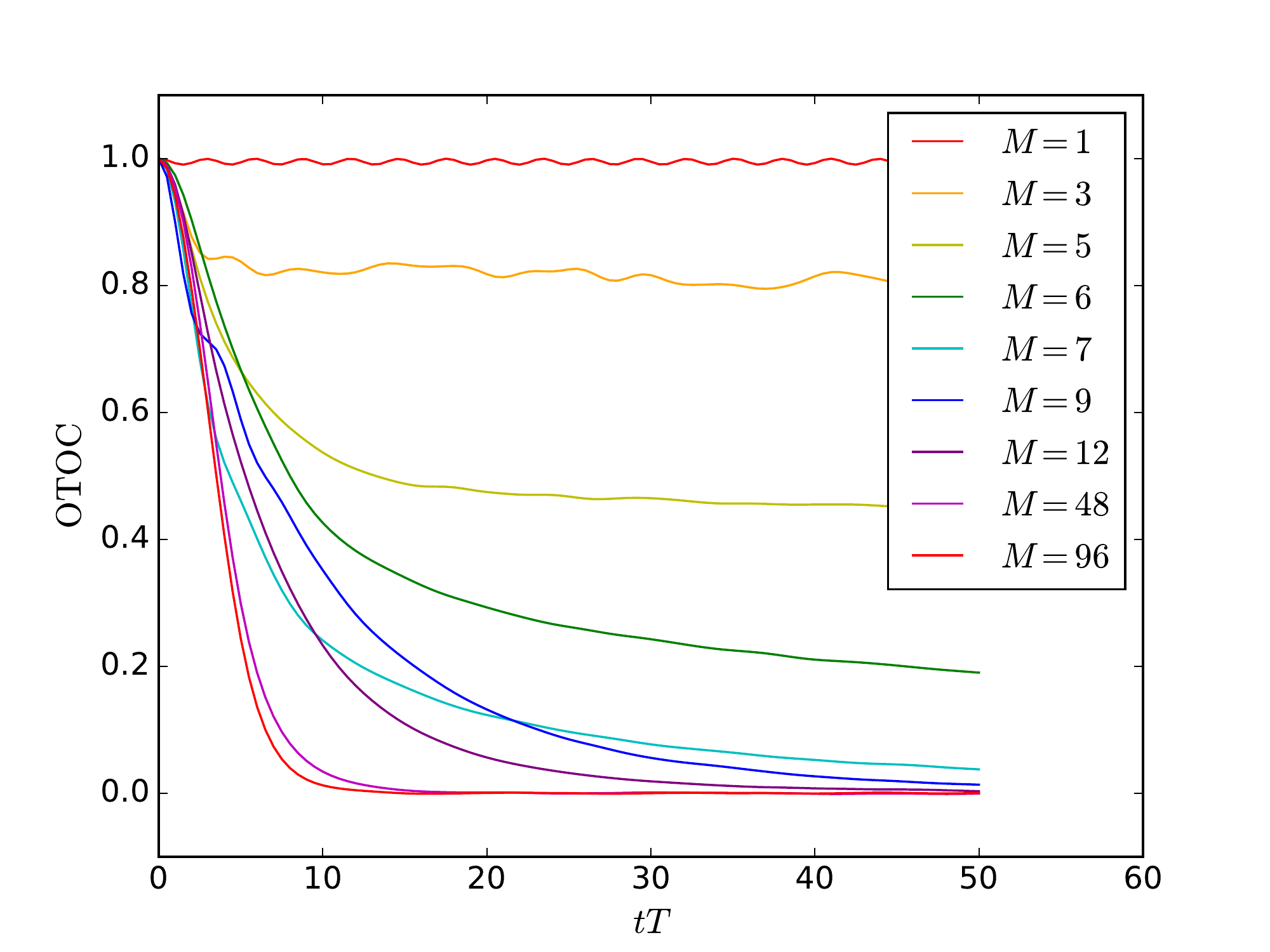}
\caption{Out-of-time-order correlator $F_{ij}(t)$ for $N=24$ with various $M$.}
\label{fig_p_OTOC}
\end{figure}

Again, the two limits of large and small $p$ show markedly different behavior in Fig.\ \ref{fig_p_OTOC}, which displays our results for $N=24$. For large $p$, the OTOC drops to zero exponentially as expected for an SYK model. For small $p$, the decay is much slower and the OTOC appears to approach a nonzero value in the long-time limit. The transition occurs around $M=6$ which corresponds to $p \approx {1\over 4}$ and agrees with our previous results for $N=30$.

Our finite-size numerics do not allow us to quantitatively capture the behavior of the OTOC expected from large-$N$ arguments, as also discussed in the recent literature [\onlinecite{Hosur2016,FuSachdev2016}]. Indeed, the Lyapunov exponent is not independent of $J$ and has only a weak dependence on temperature. Nevertheless, the transition between fast and slow scrambling behavior, expected on the basis of our heuristic arguments, is clearly visible in Fig.\ \ref{fig_p_OTOC}. 

\begin{figure*}
\centering
\includegraphics[width=0.95\textwidth]{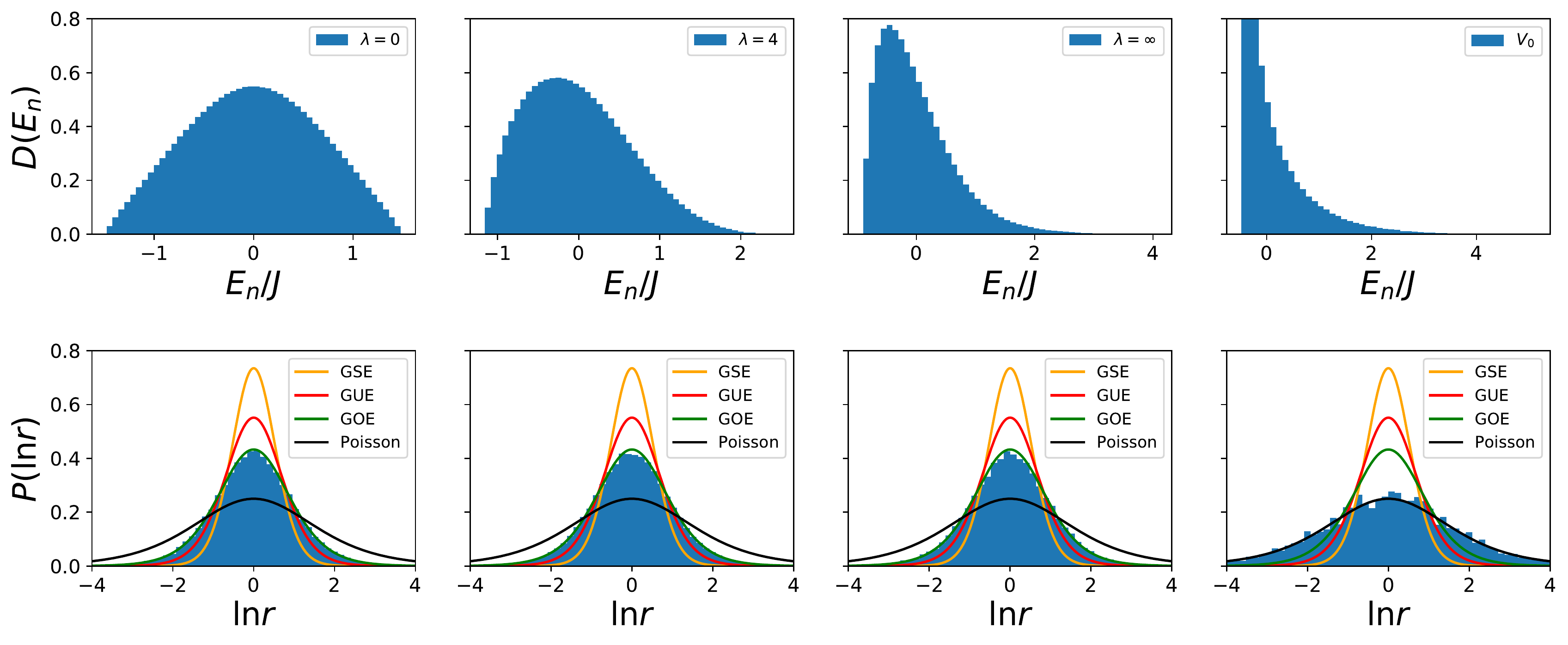}
\caption{Distribution of energy eigenvalues (top panels) and level statistics (bottom panels) for various potentials, $N=32$, $p \sim 14$ and repulsive interactions. For attractive interactions, the energy distributions are reflected around $E=0$ and level statistics remain unchanged. The potential labeled $V_0$ is the (unphysical) constant potential as described above Eq.~(\ref{eq_H4}).}
\label{fig_WF_levels}
\end{figure*}

\section{SYK$_4$ variant II: arbitrary interaction scale}
\label{sec_lambda}

In this section, we investigate the effect of an arbitrary interaction range $\lambda$ on the Hamiltonian $H_4$ defined by Eq.\ (\ref{eq_H}), while keeping $L \gg \zeta$. We are thus exploring the physics along the horizontal line at $p \gg 1$ in Fig.\ 1. In the following, we take $\zeta=1$ for simplicty.

To gain some intuition, let us first consider another limit where the model is tractable. Suppose we had an interaction potential which was \textit{constant} over the entire region where the MZMs are localized, $V_{mn} = V_0$. (This is not equivalent to $\lambda = \infty$, which corresponds to unscreened Coulomb interactions. This limit is therefore not represented in the phase diagram in Fig.~\ref{fig1}a) Then, we would have
\begin{align} 
H_4 =& \frac{V_0}{2} \left[i \sum_{i<j} \sum_m \rho_{ij}^m \chi_i \chi_j \right]^2 + E_0.
   \label{eq_H4}
\end{align}
Interestingly, the Hamiltonian reduces to the square of the \emph{non-interacting} Hamiltonian $H_2$, which was analyzed in detail in Sec.~\ref{sec_H2}, and for which we have an exact solution in the $N\rightarrow \infty$ limit. The state described by the Hamiltonian (\ref{eq_H4}) is clearly 
a dFL, and its spectrum is given by the square of the semi-circle distribution obtained for Hamiltonian $H_2$ in the limit $p \gg 1$, as shown in Fig. \ref{fig_WF_levels}.
In general, the interaction potential is neither a delta-function nor a constant and is not analytically tractable. We thus use exact diagonalization to analyze the model of Eqs.~(\ref{eq_H}) for a generic screened Coulomb potential [Eq.~(\ref{eq_Coulomb})] and system sizes up to $N=32$.

\subsection{Many-body spectra, level statistics and thermodynamics}

In Fig. \ref{fig_WF_levels}, we present the many-body spectra for the case of repulsive interactions ($u = 1$). The spectrum smoothly evolves from a symmetric distribution at small $\lambda$ (characteristic of the SYK limit analyzed in Sec.~\ref{sec_p}) towards a highly-skewed spectrum with an increased weight near the ground state. The level-statistics are remarkably robust, and follow the expected SYK statistics for all $\lambda$. By contrast, the results with constant potential $V$ follow a Poisson distribution -- a consequence of the $(H_2)^2$ structure of the Hamiltonian in Eq. (\ref{eq_H4}). For attractive interactions ($u = -1$), the spectrum is simply reflected around $E=0$, and the level statistics remain unchanged.

\begin{figure}
\centering
\includegraphics[width=0.45\textwidth]{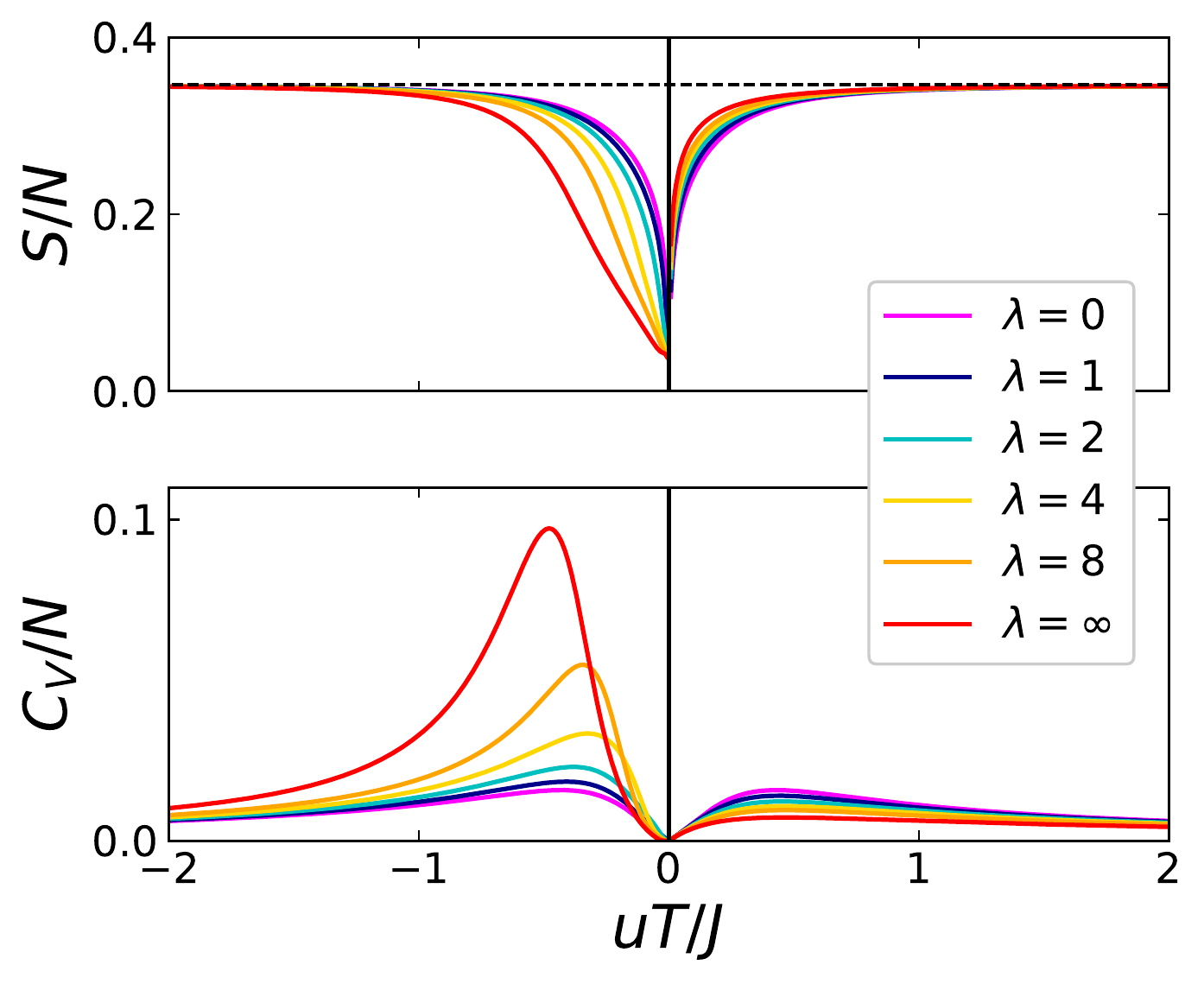}
\caption{Thermal entropy density (top) and heat capacity density (bottom) for $N=32$. Note that negative values of $uT/J$ correspond to attractive Coulomb interaction.}
\label{fig_WF_entropy}
\end{figure}

Next, we examine the thermodynamics of the model. We compute the partition function, $Z = \sum_n e^{-\beta E_n}$,
where $E_n$ are the many-body energies. The free energy and thermal entropy are obtained from
\begin{align}
    F = -T \ln Z \quad , \quad S = - \frac{\partial F}{\partial T} = \frac{ U - F}{T},
\end{align}  
where $ U = \frac{1}{Z} \sum_n E_n e^{-\beta E_n}$ is the average thermal energy. Similarly, the heat capacity is obtained as
\begin{align}
C_V =& T \frac{\partial S}{\partial T} = \frac{ \langle E_n^2\rangle - U^2}{T}.
\end{align}
We show the entropy and heat capacity densities in Fig.\ \ref{fig_WF_entropy}. Despite the finite zero-temperature entropy density predicted by the large-$N$ solution, in our finite-$N$ numerics the entropy density vanishes as $T \rightarrow 0$, in agreement with previous numerical studies [\onlinecite{FuSachdev2016}]. However, the low-temperature entropy density increases with $\lambda$ for repulsive interactions, and decreases with $\lambda$ for attractive interactions. The heat capacity shows a distinct peak building up as $\lambda$ increases for $u=-1$, which indicates a potential phase transition. The peak sharpens with increasing $N$ (not shown here), also suggesting a transition in the thermodynamic limit.

\begin{figure*}
\centering
\includegraphics[width=0.95\textwidth]{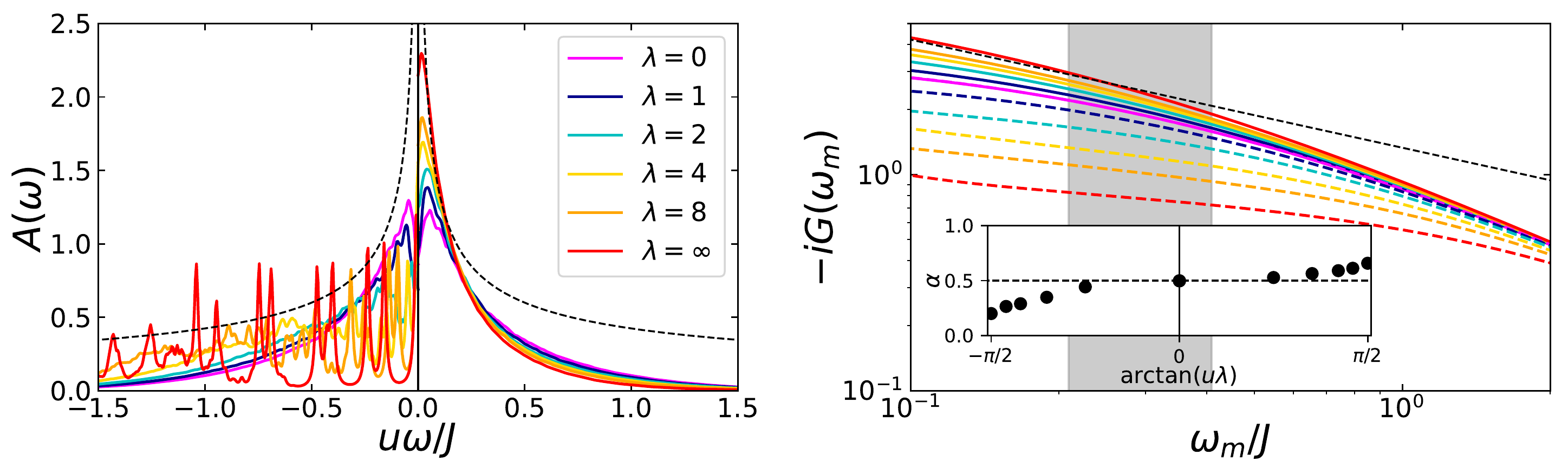}
\caption{Green's functions obtained by exact diagonalization for $N=32$, $p \sim 14$ and various values of $\lambda$. The dashed black lines indicate the conformal results for the SYK model. Left: Spectral function $A(\omega)$. Right: Matsubara Green's function $G(\omega_n)$. Solid (dashed) lines represent repulsive (attractive) interactions. Inset: Scaling exponents $\alpha$ obtained from fitting $G(\omega_n)$ to power-law behavior in the conformal regime (shown as a grey shaded area).}
\label{fig_WF_Greens}
\end{figure*}

\subsection{Green's functions}

We show the spectral function [Eq.~(\ref{eq_Aomega})] of the model  in Fig. \ref{fig_WF_Greens}, for $N=32$ and various $\lambda$. For repulsive interactions, the low-frequency divergence $\sim 1/\sqrt{\omega}$ characteristic of the SYK limit becomes stronger for increasing $\lambda$. For attractive interactions, the low-frequency divergence becomes weaker and eventually collapses, giving rise to a Fermi-liquid state with well-defined quasiparticles. We also obtain the Matsubara frequency Green's function, as shown in Fig. \ref{fig_WF_Greens}. Presumably because of finite $N$ effects, we do not observe a clean power-law behavior in the conformal regime $\frac{J}{N} \ll \omega \ll J$: the scaling exponent depends on the energy scale probed. Nevertheless, for any such scale, we obtain a clear trend as a function of $\lambda$. As an example, we show in the inset of Fig. \ref{fig_WF_Greens} the extracted scaling exponents $\alpha$ around $\omega/J = 0.3$. The exponents $\alpha$ consistently increase with $\lambda$ for repulsive interactions, and decrease for attractive interactions.
\subsection{Out-of-time-order correlators and scrambling}

\begin{figure}
\centering
\includegraphics[width = 0.45\textwidth]{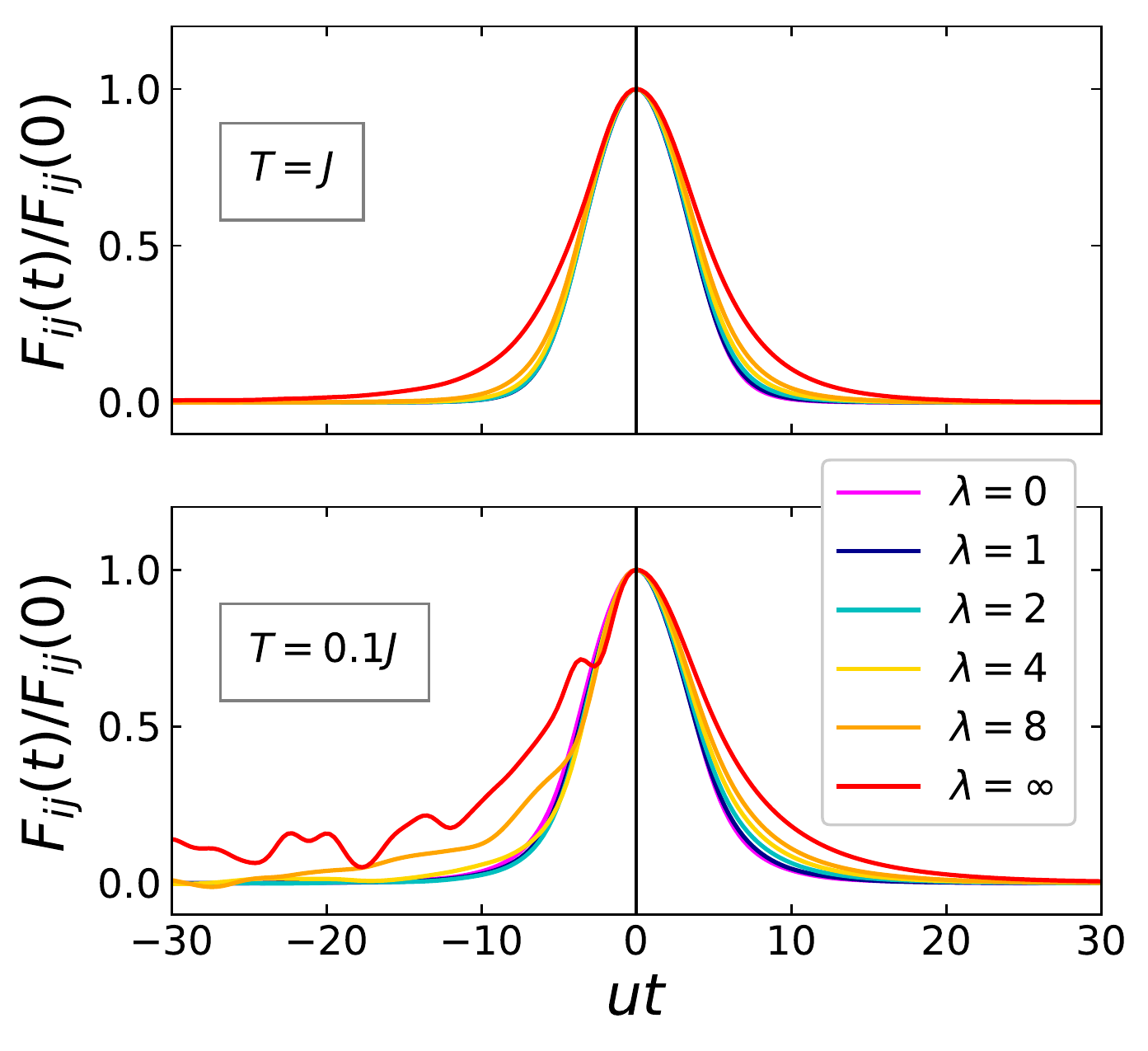}
\caption{Out-of-time-order correlators $F_{ij}(t)$ for $N=28$, $p\sim 14$ and various values of $\lambda$, at temperatures $T = J$ (top) and $T = 0.1J$ (bottom).}
\label{fig_WF_OTOC}
\end{figure}

The OTOC [Eq.~(\ref{eq_OTOC})] for our system is shown in Fig. \ref{fig_WF_OTOC}. In the high-temperature (compared to the heat capacity peak) regime $T=J$, the OTOC decays exponentially to zero in all cases. The scrambling is most efficient in the ``SYK" limit ($\lambda\to 0$) when compared to all other models in the family, which is consistent with the expectation [\onlinecite{Maldacena2016b}]. By contrast, in the low-temperature regime ($T = 0.1J$), the behavior for large $\lambda$ and attractive interactions (where the spectral function suggests a Fermi liquid) shows sub-exponential scrambling characteristic of a non-chaotic phase. This peculiar behavior, where the OTOC exhibits two different regimes depending on the temperature scale probed, is reminiscent of the transitions explored in the models of Ref.\ [\onlinecite{Bi2017,Altman2016}]. The Fermi liquid phases in these models are also strongly interacting and scramble efficiently at high temperatures. 

%

\section{Discussion}
\label{sec_Dis}

In this work, we explored three extensions of the SYK$_q$ model motivated by recent proposals for the experimental realization of this physics.
For $q=2,4$ the theory has the general structure of the SYK$_q$ model with Hamiltonians consisting of, respectively, 2-fermion and 4-fermion terms. Whereas the corresponding interaction constants $K_{ij}$ and $J_{ijkl}$ are taken as random, independent Gaussian variables in the SYK models, our extensions deviate from this assumption. Specifically, $K_{ij}$ and $J_{ijkl}$ are defined as sums with $M$ terms of products of $q$ random variables, weighted in the $q=4$ case by the screened Coulomb interaction with range $\lambda$. This specific form of the coupling constants is motivated by the proposals to realize the SYK model given in Refs.\ [\onlinecite{Pikulin2017,Alicea2017,Achen2018}]. 

The key parameter in these models is the ratio $p=M/N$. For $p\gg 1$, the coupling constants are composed of sums over a large number of identically-distributed random variables and become asymptotically Gaussian, as dictated by the central limit theorem. The models in this limit are therefore expected to exhibit SYK behavior, which we verified by analytical arguments and extensive numerical simulations. 
Lowering $p$ changes the behavior away from the canonical SYK. 

For the non-interacting system $(q=2)$, we observe a quantum phase transition at $p_c={1\over 4}$ from a gapless dFL to a gapped phase with a ground state degeneracy of $2^{(N-4M)/2}$. We obtained an exact analytical expression for the propagator $G(i\omega)$ in the large-$N$ limit and arbitrary $p$. It shows an interesting scale-invariant behavior $\sim|\omega|^{-1/3}$ at the critical point and describes both the gapped and gapless phases, which we matched to a numerical solution of the problem.

For the interacting problem $(q=4)$, the behavior depends on the range of the interaction potential $\lambda$ and its overall sign $u=\pm1$ (repulsive vs.\ attractive). For short range interactions of either sign, we find clear evidence of a phase transition at $p_c\simeq{1\over 4}$ separating the chaotic SYK-like phase from a Fermi liquid at $p<p_c$. 
The fermion scaling dimension and the scrambling properties are found to
vary continuously, interpolating between SYK at large $p$ and dFL at small $p$. Furthermore, the energy level statistics undergo a sharp transition at $p_c$ from one of the Gaussian ensembles, characteristic of the SYK model, to Poisson which is expected for a dFL.  
We have shown, using a novel large-$N$ saddle-point solution, that the model with $p\gg 1$ is indeed the SYK phase. Away from this limit, corrections become difficult to handle analytically (except for the $M=1$ case which becomes tractable again), so we relied chiefly on general arguments and numerical simulations. 

Remarkably, repulsive interactions of \emph{any} range $\lambda$ preserve the chaotic non-Fermi liquid physics found for $p \gg 1$, albeit with changed scaling dimensions and Lyapunov exponents. For attractive interactions, there appears to be a critical scale $\lambda_c$ (which might depend on $p$) above which a Fermi liquid state is recovered, as suggested by the quasiparticle peaks observed in the zero-temperature spectral function. The out-of-time order correlator, when evaluated at low temperatures compared to the characteristic energy scale $J$, then shows non-chaotic scrambling. However, chaotic scrambling is recovered at high temperatures. A prominent peak in the specific heat also hints at a thermal phase transition between the non-chaotic and chaotic regimes. Finally, and perhaps surprisingly, the energy level statistics remains robustly SYK$_4$ for all values of $\lambda$ and only becomes Poisson for constant, distance-independent interaction potential.

We therefore conjecture that the ground state for attractive, long-range interactions is a Fermi liquid but highly excited states still carry a memory of the SYK physics. That would explain why the energy level statistics (dominated by highly excited many-body states) and OTOC (evaluated at high temperature) continue to show non-Fermi liquid behavior. Clearly, more work is needed to unambiguously identify the exact nature of this state and of the phase transition described above.   

Our results have relevance to future experiments seeking to realize SYK physics in a solid-state system. They show that SYK-type non-Fermi liquids (with continuously varying fermion scaling dimension) are obtained even when the Hamiltonian deviates substantially from the ideal SYK model in several ways that can be expected to occur in a real physical system. Similar to models studied in Refs.\ [\onlinecite{Bi2017,Altman2016}], we find that the system can undergo a phase transition into a disordered Fermi liquid phase when perturbed too far from the SYK fixed point, e.g.\ for short-range interactions in the limit of small parameter $p$, or for long-range attractive interactions.  

Various interesting questions are left for future work. First, our non-interacting model (variant of SYK$_2$) exhibits a surprising scale-invariant behavior at low-energies, at the quantum critical point ($p_c = 1/4$) separating a gapped phase at low $p$ and a gapless phase at high $p$. One might wonder if there is a corresponding (0+1) conformal field theory describing that critical point, or if a higher-dimensional non-interacting model could be constructed with similar properties.
Second, we have barely scratched the finite-temperature properties of our model. This physics might be quite rich -- for example, the disordered Fermi liquids in Secs.~\ref{sec_p} and Sec.~\ref{sec_lambda} have rather different behaviors at finite temperature. The OTOC in the former case does \emph{not} decay to zero, whereas in the latter case, it eventually does at large enough temperature. Therefore, the dependence of the Lyapunov exponent on the model parameters $p$, $\lambda$, $u$ and $T$ might be non-trivial, especially near the phase transitions. Unfortunately, such an analysis is not reliable in the context of exact diagonalization and calls for more refined techniques. Finally, it is tempting to ask if our models, at least in some limit, might be mapped onto tensor-like models without disorder, similar to Witten's construction~[\onlinecite{Witten2016}].

{\em Acknowledgments --} The authors are indebted to  O\u{g}uzhan Can, Anffany Chen, and D. I. Pikulin for numerous discussions. The work described in this article was supported by NSERC and CIfAR.

\'{E}. L.-H. and C. L. contributed equally to this work.

\appendix

\section{Solution of the non-interacting Hamiltonian $H_2$}
\label{app_H2}

In this Appendix, we provide the full solution of the non-interacting variant of the SYK$_2$ model defined in Eqs. (\ref{eq_H2},\ref{eq_couplings_H2}) relevant for $\mu \gg J$,
\begin{align}
     H = \frac{i}{2} \sum_{i,j} K_{ij} \chi_i \chi_j,  \ \ \ 
    K_{ij}= -\mu \sum_{m=1}^{2M} \left( a_i^m b_j^m - a_j^m b_i^m \right) .
\end{align}
The corresponding partition function can be written as a Euclidean time path integral
\begin{widetext}
\begin{align}
    \mathcal{Z} = \int \mathcal{D}\chi e^{-S[\chi]} \quad , \quad
S[\chi] = \int d\tau \left[ \frac{1}{2} \sum_i \chi_i \partial_\tau \chi_i - i  \mu \sum_{ijm} a_i^m b_j^m \chi_i \chi_j \right].
\end{align}
We would like to perform a Hubbard-Stratonovich transformation so that we can decouple the disorder factors $(a_i^m,b_i^m)$ and average over them. To this end we introduce the following identity
\begin{align}
1 \sim \int \mathcal{D} \psi_a \mathcal{D} \psi_b \exp \left\{ - i \mu \int d\tau \sum_m \psi_a^m \psi_b^m \right\},
\end{align}
where $\psi_a$, $\psi_b$ are two flavors of Majorana fermions defined on each site $m$. The action becomes
\begin{align}
S[\psi_a, \psi_b, \chi] = \int d\tau \left[ \frac{1}{2} \sum_i \chi_i \partial_\tau \chi_i +   i \mu \sum_m \left(  \psi_a^m \psi_b^m - \sum_{ij} a_i^m b_j^m \chi_i \chi_j \right) \right].
\end{align}
Change of variables $\psi_a^m \rightarrow \psi_a^m + \sum_i a_i^m \chi_i$ and $\psi_b^m \rightarrow \psi_b^m + \sum_j b_j^m \chi_j$ then leads to
\begin{align}\label{a5}
S[\psi_a, \psi_b, \chi] = \int d\tau \left[ \frac{1}{2} \chi_i \partial_\tau \chi_i + i \mu \left(  \psi_a^m \psi_b^m -  a_i^m \psi_b^m \chi_i  + b_j^m \psi_a^m \chi_j \right)  \right]
\end{align}
where we used the fact that Grassmann  fields $\psi$, $\chi$ anticommute, and we dropped the summation symbols over repeated i. Now that we have decoupled the random wavefunction terms $a_i^m$, $b_i^m$ we can perform the average over disorder. The proper procedure for quenched disorder averages $\ln{\mathcal{Z}}$ using the replica technique, but it is known that the system is self-averaging and the same result is obtained if we compute a simple Gaussian average
\begin{align}
    \mathcal{Z}_{\rm avg} = \int \mathcal{D} a \mathcal{D} b \exp \left\{ - \frac{1}{2\sigma^2} \sum_{i,m} \left( (a_i^m)^2 + (b_i^m)^2 \right) \right\} \mathcal{Z}[a_i^m, b_i^m].
\end{align}
Here, the variance is $\sigma^2=1/8M$ based on Eq.\ (\ref{varab}).  We thus obtain an effective action
\begin{align}
S[\psi_a, \psi_b, \chi] = \frac{1}{2\sigma^2} \sum_{i,m} \left( (a_i^m)^2 + (b_i^m)^2 \right) + \int d\tau \left[ \frac{1}{2} \chi_i \partial_\tau \chi_i + i \mu\left(  \psi_a^m \psi_b^m - a_i^m \psi_b^m \chi_i  + b_i^m \psi_a^m \chi_i \right)  \right].
\end{align}
Completing the square and performing the integral over $a$, $b$ we find
\begin{align}
S[\psi_a, \psi_b, \chi] = \int d\tau \left[ \frac{1}{2} \chi_i \partial_\tau \chi_i +  i \mu  \psi_a^m \psi_b^m \right] - \frac{\mu^2 \sigma^2}{2} \int d\tau d\tau' \left[\psi_b^m(\tau) \psi_b^m(\tau') \chi_i(\tau) \chi_i(\tau') \right]   +  ( a \leftrightarrow b). 
\end{align}
In order to solve this action in the saddle-point approximation, we define the Green's functions
\begin{align}
    G(\tau, \tau') = \frac{1}{N} \sum_{i=1}^N \chi_i(\tau) \chi_i(\tau') \quad , \quad F_{a,b}(\tau, \tau') = \frac{1}{2M} \sum_{m=1}^{2M} \psi_{a,b}^m(\tau) \psi_{a,b}^m(\tau'),
\end{align}
by means of inserting the following integrals into the action
\begin{align}
1 &\sim \int D\Sigma DG \exp \left \{ - \frac{N}{2} \Sigma(\tau, \tau') \left( G(\tau, \tau') - \frac{1}{N} \sum_i \chi_i(\tau) \chi_i(\tau') \right) \right\}, \\
1 &\sim \int D\Omega_a DF_a \exp \left \{ - M \Omega_a(\tau, \tau') \left( F_a(\tau, \tau') - \frac{1}{2M} \sum_m \psi_a^m(\tau) \psi_a^m(\tau') \right) \right\} \quad , \quad (a \leftrightarrow b),
\end{align}
where $\Sigma$, $\Omega_{a/b}$  are self-energies that serve as Lagrange multipliers. The action thus becomes
\begin{align}
S[\psi_a, \psi_b, \chi] =& \int d\tau d\tau' \Bigg[ \frac{1}{2} \chi_i \left( \delta(\tau-\tau') \partial_\tau  - \Sigma  \right) \chi_i + i \mu  \psi_a^m \psi_b^m - \frac{1}{2} \Omega_a \psi_a^m \psi_a^m - \frac{1}{2} \Omega_b \psi_b^m \psi_b^m  \nonumber \\
&  - \mu^2 \sigma^2 NM\left[F_a + F_b \right] G   + M (F_a \Omega_a + F_b \Omega_b) + \frac{1}{2} N \Sigma G \Bigg]
\end{align}
We now use the relation $M = pN$, and introduce the energy scale $K$ previously defined by Eq.\ (\ref{K}) as  $K^2=\mu^2 / 16p = \mu^2 \sigma^2 N/2$ to rewrite the action as
\begin{align}
S[\psi_a, \psi_b, \chi] =& \int d\tau d\tau' \Bigg[ \frac{1}{2} \chi_i \left( \delta(\tau-\tau') \partial_\tau  - \Sigma  \right) \chi_i + \frac{1}{2} \left(\psi_a^m, \psi_b^m \right) 
\begin{pmatrix}
-\Omega_a & i \mu\\
-i \mu & -\Omega_b 
\end{pmatrix}
\begin{pmatrix}
\psi_a^m \\
\psi_b^m
\end{pmatrix} \nonumber \\
&  - 2 K^2 N p \left[F_a + F_b \right] G   + pN (F_a \Omega_a + F_b \Omega_b) + \frac{1}{2} N \Sigma G \Bigg].
\end{align}
\end{widetext}
Written in this form, we can now deduce the saddle-point equations. This is most easily done by first passing into Matsubara frequency representation, where one can easily perform the required integrals over Grassmann fields. We thus obtain 
\begin{align}
\frac{ \partial S_0}{\partial \Sigma} &= G(i\omega) -\frac{1}{- i\omega - \Sigma(i \omega)} = 0 \ \ , \\
\frac{ \partial S_0}{\partial \Omega_a} &= p F_a(i\omega) + p \frac{\Omega_b(i\omega)}{\Omega_a(i\omega) \Omega_b(i\omega) - \mu^2} = 0 \ \  (a \leftrightarrow b), \\
\frac{ \partial S_0}{\partial G} &= \frac{1}{2}\Sigma(i\omega) - 2 K^2 p \left[F_a(i\omega) + F_b(i\omega) \right] = 0 \ \ , \\
\frac{ \partial S_0}{\partial F_a} &= p \Omega_a(i\omega) - 2 p K^2 G(i\omega) = 0 \ \ (a \leftrightarrow b) .
\end{align}
In the following we will systematically suppress the frequency dependence for the sake of brevity. It is clear that $\Omega_a = \Omega_b$ at the saddle point, which in turn indicates that $F_a = F_b$. So the saddle-point equations simplify
\begin{align}
G = \frac{1}{- i\omega - \Sigma}, \ \ \ \ \Sigma = 8 p K^2  F  \\
F = \frac{ \Omega}{16 p K^2 - \Omega^2}, \ \
\ \
\Omega = 2 K^2 G.
\end{align}
Solving this system of equations yields a single cubic equation for the Green's function of interest
 \begin{align}\label{cub1}
 i \omega K^2 G^3 + (1 - 4p)K^2 G^2 - 4 i\omega p G - 4p  = 0.
\end{align}

The cubic equation (\ref{cub1}) can be solved analytically for $G(i\omega)$ using the Cardano formula, but the expression is lengthy and does not provide any useful insights. It is instructive, however, to solve Eq.\ (\ref{cub1}) in various limits.
When $p\gg 1$, we can neglect the cubic term, leading to
$ K^2 G^2 + i\omega G + 1  = 0$.
Continuing to real frequencies using $-i\omega_n \rightarrow \omega+i\delta$, we obtain
 \begin{align}
 G(\omega) = \frac{\omega}{2 K^2} \pm \frac{i}{K} \sqrt{1 - \left( \frac{\omega}{2K}\right)^2 }
\end{align}
which leads to the semi-circle law for the spectral function, Eq. (\ref{semi}). For $p\ll 1$, we can neglect the last two terms in Eq.\ (\ref{cub1}) which gives $G\simeq -(1-4p)/i\omega$. Upon analytical continuation this leads to a low-frequency spectral function $A(\omega)\simeq (1-4p)\delta(\omega)$ reflecting the expected manifold of degenerate zero modes in this limit. With a modest amount of additional work, one can also deduce the lobes in $A(\omega)$ centered around $\omega=\pm K/2\sqrt{p}$ that are expected on the basis of arguments given in Sec.\ III. At the critical point, $p = 1/4$, we obtain:
 \begin{align}
 i \omega K^2 G^3 -  i\omega G - 1  = 0.
 \label{eq_criticalpoint_G}
\end{align}
It is straightforward to show that, for $\omega \rightarrow 0$, the second term of Eq. (\ref{eq_criticalpoint_G}) can be neglected, leading to a power-law solution for the spectral function, $A(\omega) \sim |\omega|^{-1/3}$.

\section{Solution to the model in Section~\ref{sec_p}}\label{App-4phi}
The Hamiltonian of the model is 
\begin{align}\label{hh1}
H&=\sum_{i<j<k<l}^N J_{ijkl}\chi_i\chi_j\chi_k\chi_l, \nonumber \\
J_{ijkl} &=-\frac{V_0}{3\zeta}\sum_{m=1}^M\sum_{\alpha\beta\mu\gamma=1}^4 \epsilon^{\alpha\beta\mu\gamma}\phi_{\alpha i}^m\phi_{\beta j}^m\phi_{\mu k}^m\phi_{\gamma l}^m
\end{align}
where $\phi$'s are Gaussian-distributed random real numbers with 
\begin{equation}
\overline{\phi_{\alpha i}^m}=0,\ \ \ \ 
\overline{\phi_{\alpha i}^m\phi_{\beta j}^{m'}}=\frac{1}{M}\delta_{\alpha\beta}\delta_{ij}\delta^{mm'}.
\end{equation}
Using the anti-commutation relations of $\chi$'s, the Hamiltonian can be rewritten as
\begin{align}\label{hh3}
H&=\frac{u}{2}\sum_m\sum_{\substack{i,j,k,l\\ \mathrm{all\ different}}}
\phi_{1i}^m\phi_{2j}^m\phi_{3k}^m\phi_{4l}^m\chi_i\chi_j\chi_k\chi_l  \\
&=\frac{u}{2}\sum_m\sum_{\substack{i,j,k,l\\ \mathrm{all\ different}}}
(\phi_{1i}^m\chi_i)(\phi_{2j}^m\chi_j)(\phi_{3k}^m\chi_k)(\phi_{4l}^m\chi_l), \nonumber
\end{align}
where we introduced $u=-2V_0/3\zeta$ to simplify notation. To proceed, we want to average over the random variables $\phi$. We thus recast the problem defined by Hamiltonian (\ref{hh3}) as an imaginary time path integral with the Euclidean action $S[\chi]
=\int\mathrm{d}\tau(\frac{1}{2} \sum_j \chi_j\partial_{\tau}\chi_j+H)$
and use an identity for Grassmann variables
\begin{equation}\label{hh4}
e^{-\frac{1}{g}\eta_1\eta_2\eta_3\eta_4}
=\int\mathcal{D}\psi e^{-g\psi_1\psi_2\psi_3\psi_4-i\sum_{\alpha=1}^4\psi_{\alpha}\eta_{\alpha}},
\end{equation}
to decouple the disorder terms. The identity can be checked by expanding the exponential under the integral and performing the integration term by term. 
\begin{widetext}
Using this identity with  $\eta_{\alpha}^m=\sum_i\phi_{\alpha i}^m\chi_i$, we obtain $e^{-S[\chi]}=\int\mathcal{D}\psi e^{-S_{\mathrm{eff}}[\chi,\psi]}$ with 
\begin{equation}
S_{\mathrm{eff}}[\chi,\psi]
=\int\mathrm{d}\tau\Big(\frac{1}{2}\sum_j\chi_j\partial_{\tau}\chi_j
+\frac{2}{u}\sum_m\psi_1^m\psi_2^m\psi_3^m\psi_4^m
+i\sum_{m,\alpha,j}\psi_{\alpha}^m\phi_{\alpha j}^m\chi_j\Big)
\end{equation}
Now the partition function can be averaged over disorder (subject to the caveat noted below Eq. \ref{a5}). Using 
\begin{equation}
\mathcal{Z}\rightarrow \int \mathcal{D}\phi \exp\Big(-\frac{M}{2}\sum_{\alpha,i,m}(\phi_{\alpha i}^m)^2\Big)\mathcal{Z}
\end{equation}
we find
\begin{equation}
S_{\mathrm{eff}}[\chi,\psi]
=\int\mathrm{d}\tau\Big(\frac{1}{2} \sum_j \chi_j\partial_{\tau}\chi_j
+\frac{2}{u}\sum_m\psi_1^m\psi_2^m\psi_3^m\psi_4^m\Big)+\frac{1}{2M}\int\mathrm{d}\tau_1\mathrm{d}\tau_2
\sum_{m,\alpha}\big(\psi_{\alpha}^m(\tau_1)\psi_{\alpha}^m(\tau_2)\big)\sum_j\big(\chi_j(\tau_1)\chi_j(\tau_2)\big).
\end{equation}
We next define $G(\tau_1,\tau_2)=\frac{1}{N}\sum_j\chi_j(\tau_1)\chi_j(\tau_2)$ by inserting
\begin{equation}
1=\int\mathcal{D}G\mathcal{D}\Sigma\exp\bigg(-\int\mathrm{d}\tau_1\mathrm{d}\tau_2
\frac{N}{2}\Sigma(\tau_1,\tau_2)\Big(G(\tau_1,\tau_2)-\frac{1}{N}\sum_j\chi(\tau_1)\chi(\tau_2)\Big)\bigg)
\end{equation}
into the partition function. Integrating over the $\chi_j$ fields we finally obtain
\begin{equation}\label{B9}
\begin{split}
S_{\mathrm{eff}}[\psi,G,\Sigma]
&=N\sum_{\omega>0}\ln(-i\omega-\Sigma)+\frac{2}{u}\int\mathrm{d}\tau\sum_m\psi_1^m\psi_2^m\psi_3^m\psi_4^m\\
&\quad+\int\mathrm{d}\tau_1\mathrm{d}\tau_2\Big(\frac{N}{2}\Sigma(\tau_1,\tau_2) G(\tau_1,\tau_2)
+\frac{N}{2M}G(\tau_1,\tau_2)
\sum_{m,\alpha}\big(\psi_{\alpha}^m(\tau_1)\psi_{\alpha}^m(\tau_2)\big)\Big).
\end{split}
\end{equation}

The problem has been simplified considerably.
If we were able to integrate out the auxiliary fermions $\psi$, we could employ the large-$N$ approximation and solve the problem through the saddle-point equations. At the saddle point, $G(\tau_1,\tau_2)=G(\tau_1-\tau_2)$ is a classical function and the action for each flavor $m$ of the $\psi$ fermion is the same and given by 
\begin{equation}\label{B10}
e^{-S_4}=\int\mathcal{D}\psi \exp\Big(-\frac{2}{u}\int\mathrm{d}\tau\psi_1\psi_2\psi_3\psi_4
-\frac{N}{2M}\int\mathrm{d}\tau_1\mathrm{d}\tau_2G(\tau_1,\tau_2)
\sum_{\alpha}\big(\psi_{\alpha}(\tau_1)\psi_{\alpha}(\tau_2)\big)\Big).
\end{equation}
The action can thus be written as 
\end{widetext}
\begin{equation}\label{B11}
S_{\mathrm{eff}}[\psi,G,\Sigma]
=N\sum_{\omega>0}\left[\ln(-i\omega-\Sigma)+{1\over 2}\Sigma G\right]-MS_4.
\end{equation}
and the corresponding saddle point equations read
\begin{equation}
\begin{split}
&\frac{\delta S_{\mathrm{eff}}}{\delta\Sigma}=0\ \Rightarrow \  G^{-1}(i\omega)=-i\omega-\Sigma(i\omega),\\
&\frac{\delta S_{\mathrm{eff}}}{\delta G}=0\ \Rightarrow \ \Sigma(\tau_1,\tau_2)=-\frac{\delta S_4}{\delta G(\tau_1,\tau_2)}.
\end{split}
\end{equation}
Also note that
\begin{equation}
\frac{\delta S_4}{\delta G(\tau_1,\tau_2)}=\frac{N}{2M}\Big\langle\sum_{\alpha}\mathcal{T}_{\tau}\psi_{\alpha}(\tau_1)\psi_{\alpha}(\tau_2)\Big\rangle_{S_4}.
\end{equation}
The saddle point equations thus read
\begin{equation}\label{saddle}
\begin{split}
&G^{-1}(i\omega)=-i\omega-\Sigma(i\omega),\\
&\Sigma(\tau_1,\tau_2)=-\frac{1}{2p}\Big\langle\sum_{\alpha}\mathcal{T}_{\tau}\psi_{\alpha}(\tau_1)\psi_{\alpha}(\tau_2)\Big\rangle_{S_4},
\end{split}
\end{equation}
where the last expectation value is to be evaluated with respect to action $S_4$. 

The action $S_4$ defined through Eq.\ (\ref{B10}) appears simple as it involves only 4 Majorana fermions. Despite its apparent simplicity, we were unable to evaluate it for a general (unknown) function $G(\tau_1,\tau_2)$. As a result we do not have a closed-form solution to the problem valid in the $N\to \infty$ limit.
In the following we will solve the problem to leading order at large $p$. 

To make notation clearer, we rewrite the action (\ref{B11}) as $S_{\rm eff}=S_{\mathrm{cl}}[f]+MS'[\psi]$ with
\begin{align}
S'[\psi]=&\int\mathrm{d}\tau_1\mathrm{d}\tau_2f(\tau_1-\tau_2)\sum_{j=1}^4\psi_j(\tau_1)\psi_j(\tau_2) \nonumber \\
&+g\int\mathrm{d}\tau\psi_1(\tau)\psi_2(\tau)\psi_3(\tau)\psi_4(\tau)
\end{align}
where we define
\begin{equation}
f(\tau_1-\tau_2)=\frac{1}{2p}G(\tau_1-\tau_2),\ g=\frac{2}{u}
\end{equation}
Now, the fact that the functional form of $f$ is unknown obstructs the usual perturbative approach (expanding in powers of $g$) where we have to deal with Matsubara sums such as
\begin{equation}
\sum_{\lbrace\omega\rbrace} \frac{1}{f},\ \sum_{\lbrace\omega\rbrace} \frac{1}{fff},\ \mathrm{etc.}
\end{equation}
Thus we turn to the opposite limit, where the $f$ term serves as perturbation. This approach is expected to be valid for large $p$ and in the long-time limit, since we expect $G(\tau)$ to decay as $\tau\to \infty$.

The quantity we want to calculate is
\begin{equation}
\langle\mathcal{T}_{\tau}\psi_1(\tau_1)\psi_1(\tau_2)\rangle
=\frac{\int \mathcal{D}\psi \psi_1(\tau_1)\psi_1(\tau_2)e^{-S'[f]}e^{-S'_g}}{\int \mathcal{D}\psi e^{-S'[f]}e^{-S'_g}}
\end{equation}
In order for the numerator to be nonzero, we need a term proportional to
\begin{equation}
\psi_2(\tau_1)\psi_2(\tau_2)\psi_3(\tau_1)\psi_3(\tau_2)\psi_4(\tau_1)\psi_4(\tau_2)
\end{equation}
from $e^{-S'[f]}$. Thus we expand $e^{-S'[f]}$ up to $f^3$, and the numerator at this order is
\begin{equation}
(-2f(\tau_1-\tau_2))^3(-g)^{\mathcal{N}-2}
\end{equation} 
where the factor $2$ comes from exchanging $\tau_1\leftrightarrow\tau_2$ and $\mathcal{N}$ is the discrete path integral step. On the other hand, for the denominator, the result remains unchanged for an $f^3$ expansion, that is
\begin{equation}
(-g)^{\mathcal{N}}
\end{equation}
The final result is thus
\begin{equation}
\langle\mathcal{T}_{\tau}\psi_1(\tau_1)\psi_1(\tau_2)\rangle
=-\frac{8}{g^2}f^3(\tau_1-\tau_2)=-\frac{u^2}{4p^3}G^3(\tau_1-\tau_2)
\end{equation}
which is further substituted to the saddle point equation
\begin{equation}
\Sigma(\tau_1-\tau_2)=-\frac{1}{2p}\Big\langle\sum_{\alpha}\mathcal{T}_{\tau}\psi_{\alpha}(\tau_1)\psi_{\alpha}(\tau_2)\Big\rangle=\frac{u^2}{2p^4}G^3(\tau_1-\tau_2).
\end{equation}
Together with
$G^{-1}(i\omega)=-i\omega-\Sigma(i\omega)$
this is seen to coincide with the SYK saddle point result, provided that we identify $J^2=u^2/2p^4$.

\bibliography{SYK}

\end{document}